\documentclass[pre,a4paper,superscriptaddress,showpacs,nofootinbib]{revtex4}
 
\usepackage{amsmath}
\usepackage{amsfonts}
\usepackage{amssymb}
\usepackage{bm}
\usepackage{latexsym}
\usepackage{isolatin1}

\usepackage{graphicx}

\usepackage{float}

\newcommand{\Rend}{\mbox{$R_\text{e}$}}
\newcommand{\Rgyr}{\mbox{$R_\text{g}$}}
\newcommand{\Ree}{\mbox{$\left< R_\text{e}^2 \right>$}}
\newcommand{\Rgg}{\mbox{$\left< R_\text{g}^2 \right>$}}
\newcommand{\phistar}{\mbox{$\phi^*$}}

\newcommand{\kB}{k_{\text{B}}}

\newcommand{\dend}{\mbox{$d_{\text{end}}$}}

\newcommand{\Ne}{N_\text{e}}
\newcommand{\de}{\mbox{$d_\text{e}$}}
\newcommand{\Ds}{\mbox{$D_\text{s}$}}
\newcommand{\Dc}{\mbox{$D_\text{c}$}}
\newcommand{\taus}{\mbox{$\tau_\text{s}$}}
\newcommand{\tauc}{\mbox{$\tau_\text{c}$}}

\newcommand{\tauone}{\mbox{$\tau_1$}}
\newcommand{\tautwo}{\mbox{$\tau_2$}}
\newcommand{\tauthree}{\mbox{$\tau_3$}}
\newcommand{\taufour}{\mbox{$\tau_4$}}
\newcommand{\tausix}{\mbox{$\tau_{\text{fp}}$}}
\newcommand{\tauseven}{\mbox{$\tau_{N/2}$}}

\newcommand{\taut}{\mbox{$\tau_\text{t}$}}
\newcommand{\taui}{\mbox{$\tau_\text{i}$}}
\newcommand{\tauend}{\mbox{$\tau_{\text{end}}$}}
\newcommand{\taud}{\mbox{$\tau_{d}$}}

\newcommand{\splus}{\mbox{$s_+$}}
\newcommand{\tauplus}{\mbox{$\tau_+$}}
\newcommand{\tauminus}{\mbox{$\tau_-$}}
\newcommand{\taucoup}{\mbox{$\tau_{\text{coup}}$}}
\newcommand{\taufluc}{\mbox{$\tau_{\text{fluc}}$}}

\newcommand{\mminus}{\mbox{$m$}}
\newcommand{\mplus}{\mbox{$m_+$}}

\newcommand{\gi}{\mbox{$g_{i}$}}

\newcommand{\gone}{\mbox{$g_{1}$}}
\newcommand{\gtwo}{\mbox{$g_{2}$}}
\newcommand{\gthree}{\mbox{$g_{3}$}}
\newcommand{\gfour}{\mbox{$g_{4}$}}
\newcommand{\gfive}{\mbox{$g_{5}$}}
\newcommand{\gsix}{\mbox{$g_{6}$}}

\newcommand{\rsfig}[1]{\begin{center}
                       \includegraphics*[width=0.5\textwidth]{{#1}}
                       \end{center}
                       }

\begin{document}
\sloppy

\title{Dynamical Properties of the Slithering Snake Algorithm:\\ 
A numerical test of the activated reptation hypothesis}

\author{L.~Mattioni}
\affiliation{D\'epartement de Physique des Mat\'eriaux, Universit\'e Claude Bernard \& CNRS, 69622 Villeurbanne Cedex, France}

\author{J.~P.~Wittmer}
\email[Email: ]{jwittmer@dpm.univ-lyon1.fr}
\affiliation{D\'epartement de Physique des Mat\'eriaux, Universit\'e Claude Bernard \& CNRS, 69622 Villeurbanne Cedex, France}

\author{J. Baschnagel}
\affiliation{Institut Charles Sadron, 6 rue Boussingault, 67083 Strasbourg,
France}

\author{J.-L. Barrat}
\affiliation{D\'epartement de Physique des Mat\'eriaux, Universit\'e Claude Bernard \& CNRS, 69622 Villeurbanne Cedex, France}

\author{E.~Luijten}
\altaffiliation[Current address: ]{University of Illinois at Urbana-Champaign}
\affiliation{Institut f\"ur Physik, Johannes Gutenberg-Universit\"at Mainz, 55099 Mainz, Germany}
\affiliation{Dept. of Materials Science and Engineering, University of Illinois at Urbana-Champaign, 1304 W.~Green St., Urbana, IL 61801, USA} 

\begin{abstract}
%
%
Correlations in the motion of reptating polymers in a melt
are investigated by means of Monte Carlo simulations of the
three dimensional slithering snake version of the bond-fluctuation model.
%
%
Surprisingly, the slithering snake dynamics becomes inconsistent with 
classical reptation predictions at high chain overlap
(created either by chain length $N$ or by the volume fraction $\phi$
of occupied lattice sites), 
where the relaxation times increase much faster than expected. 
This is due to the anomalous curvilinear diffusion
in a finite time window whose upper bound 
$\tauplus(N)$ \ is set by the density of chain ends $\phi/N$. 
Density fluctuations created by passing chain ends allow a reference
polymer to break out of the local cage of immobile obstacles
created by neighboring chains.
The dynamics of dense solutions of ``snakes'' at $t \ll \tauplus$
is identical to that of a benchmark system where all chains but one 
are frozen.
%
We demonstrate that the subdiffusive dynamical regime is caused by the slow 
creeping of a chain out of its
correlation hole. 
Our results are in good qualitative agreement with the activated reptation
scheme proposed recently by Semenov and Rubinstein
[Eur.\ Phys.\ J. B, {\bf 1} (1998) 87].
Additionally, we briefly comment on the relevance of local relaxation pathways
within a slithering snake scheme.  Our preliminary results suggest that a 
judicious choice of the ratio of local to slithering snake moves is crucial to 
equilibrate a melt of long chains efficiently. 
\end{abstract}



\maketitle

\section{Introduction}
\label{sec:intro}

\paragraph*{Background.}
The molecular-level description of the dynamics of non-dilute polymer
liquids remains a challenging problem of great fundamental interest
as well as practical importance \cite{LRP90}.
In the last 30 years, attention has largely focused on the applicability
of the {\em reptation} concept introduced by de Gennes and Edwards
\cite{DG79,DE86}.
This approach postulates that individual chains, constrained by their neighbors,
move primarily along their own contours in a snakelike fashion.
Reptation-based models have been very successful in reconciling a wide range
of experimental observations \cite{LRP90} and are broadly in agreement with
more recent computational results as reviewed in \cite{KP97,KG95}. 
All evidence amounts to the existence of an additional and chain length independent
length scale. This length scale is called the ``tube diameter" $\de \sim \Ne^{\nu}$ 
with $\Ne(\phi,l_\text{p})$ the associated ``entanglement length",
$\phi$ the molecular density,
$l_p$ the persistence length
and $\nu$ the Flory exponent of the polymer chain.
The density and persistence length dependence of $\de$ have been 
probed by various numerical methods \cite{BB00,PBHK91,PBHK91_JCP,MWB00,FM01},
indicating that it is proportional to the excluded volume blob size $\xi(\phi,l_\text{p})$.
While, strictly speaking, not having the (more prestigious) status of a 
`scientific theory', the reptation concept is widely (though not generally) 
seen as a physically appealing starting point any further modeling approach 
has to compete with.

\paragraph*{First motivation.}
The simulation of entangled polymer melts is a challenging task 
and it is a matter of debate if the (suspected) reptating motion of
polymer chains has already been reproduced numerically in a
sufficiently large simulation box \cite{PKG00}. 
Without any doubt, it will be near to impossible to simulate within 
the near future by means of {\em locally} realistic dynamics 
industrially relevant problems,
such as the rheology of polymer melts with fillers.
It is, therefore, tempting to accept the snakelike motion
as a basic ingredient in a coarse-grained algorithm and to
verify if a meaningful dynamic interpretation of the well-known 
``slithering snake algorithm" (SSA) \cite{Kron,Wall,Sokal} can be given.
This algorithm is sketched in Fig.~\ref{figsketchAlgo}.
It has been extensively used for three decades in simulations of
static properties because of its efficiency in exploring phase space. 
Clearly, it would be extremely useful if a (simple and transparent) 
{\em mapping} of this algorithm on its corresponding 
local dynamical scheme could be achieved.

\paragraph*{Self-consistency of reptation.}
While most of the current computational work appears to focus on the
demonstration of the curvilinear chain motion \cite{KG95,MWB00,FM01,KBM01,PKG00},
here we take this key reptation hypothesis for granted. 
Therefore, we are not interested in the Rouse-like dynamics at short times
and for chain length smaller than $\Ne$. 
For the SSA we expect $\Ne \approx 1$ so that it should be possible to 
explore the dynamics of the so-called `primitive path' \cite{DE86}.
What we do question, however, is an additional (and arguably less profound)  
mean-field assumption made in the original versions of reptation 
which supposes the {\em uncorrelated} motion of neighboring ``snakes''.%
\footnote{We will often use the term ``snake'' as a short form for
``chains moving according to the slithering snake algorithm''.}
In the following we refer to this assumption as the ``uncorrelated reptation 
hypothesis" (URH).
One of its immediate consequences is, for instance, the linear time dependence of
the mean-square displacements (MSD) of the center of mass or of the curvilinear 
motion of an inner chain monomer for the snake dynamics.
The verification of the influence of the 
interchain correlations on reptating snakes is the second 
motivation of this study.

\paragraph*{Internal modes, tube fluctuations.}
In recent years a number of theoretical models and numerical schemes 
\cite{DE86,Doi83,Rubinstein87,DesCloizeaux} 
have been developed in order to explain the observed molecular weight dependence of the 
zero-shear viscosity $\eta \sim N^{3.4 \pm 0.2}$ for $N \gg N_e$ \cite{LRP90,DE86}. 
The main idea of these theories is to account for the internal modes of polymer dynamics 
inside the reptation tube (``tube-length fluctuations") or to introduce a finite lifetime 
of the tube (``double reptation" \cite{DesCloizeaux}) 
without addressing, however, the possibility of correlated chain motion. 
While these are undoubtedly valiant efforts, they are of no concern here,
as we are, to stress it again, interested in the physics at much larger
molecular weights and correspondingly longer time scales.

\paragraph*{Coupling between dynamics and density fluctuations.}
The possibility of a coupling between primitive path dynamics and the fluctuating excluded 
volume interactions of neighboring chains was first investigated by Deutsch \cite{Deutsch}. 
He suggested that the reptation of a chain tail from its original tube to a new
environment implies a finite free energy penalty $F(s)$ proportional to the 
curvilinear length of the tail $s$. 
Moreover, it was assumed that this additional activation energy of
the snake can relax as soon as the spatial size of the escaped portion becomes
of the order of the mean distance between chain ends $\dend \sim N^{1/3}$. Taking into
account the Gaussian chain statistics in a dense polymer solution,
Deutsch obtained an exponential increase 
$\tau \sim \eta \sim N^3 \exp(N^{2/3})$ for the relaxation times and viscosities 
in concentrated polymer solutions with extremely large molecular masses.
More recently, these ideas have been elaborated further by Semenov and 
Rubinstein 
\cite{Semenov91,Semenov97,SR98} and implemented in a slightly different
and thermodynamically fully consistent way.
They predict that the URH breaks down for $N \gg \Ne^3$ \cite{SR98}.
Since we expect $\Ne$ to be of order 1 for the SSA, it should be possible to 
test this ``{\em activated reptation}" hypothesis (ARH) in our simulations.

\paragraph*{Brief summary of our work.}
%
We have harnessed a SSA of the bond-fluctuation lattice model and compared its 
dynamics for consistency with the URH of the classical reptation model. 
We find that the dynamics rapidly slows down with chain overlap
and becomes strongly subdiffusive in a finite time interval
$\tauminus \ll t \ll \tauplus$ whose bounds are determined by the blob size $\xi$
and the mean distance between chain ends $\dend$.
This slowing down is at variance with the URH and indicates strong
correlations between neighboring chains.
To corroborate our results for the SSA dynamics in an annealed molecular field, 
referred to as ``annealed SSA'' (aSSA) hereafter,
we have considered a different version of the SSA where all chains but one have been 
quenched. We will refer to this version as the ``quenched SSA" (qSSA).
The comparison of both dynamical schemes demonstrates the mutual influence of
the correlation hole retaining the snakes by means of an effective force
and the fluctuating chain ends which allow the snakes to sneak out of the
local cage of immobile monomer sites. 

\paragraph*{Outline.} 
This work is presented as follows.
Our numerical scheme is briefly recapitulated in the next section.
The parameter range and some technical points concerning the configurations
and measurements are presented in Sec.~\ref{sec:config}.
%
%
The relevant analytical predictions according to the URH and ARH
are summarized in Sec.~\ref{sec:theory}.
The numerical results of this paper are presented in Sec.~\ref{sec:results}
where we characterize and compare aSSA and qSSA dynamics.
The implications and possible extensions of our work are discussed 
in the final Sec.~\ref{sec:conclusion}.
There, we speculate on the importance of local motion to obtain
a fully consistent mapping of a slithering snake scheme 
on a locally realistic model.

\section{Simulation model}
\label{sec:simulation}

\paragraph*{Bond-fluctuation model.}
In this investigation we use the extensively studied `bond-fluctuation model' (BFM)
\cite{PBHK91,DB91,Marcus},
a lattice Monte Carlo scheme introduced by Carmesin and Kremer \cite{CK88}.
Figure~\ref{figsketchAlgo} depicts a two dimensional projection of
both the original BFM with local monomer dynamics (upper chain)
and the BFM with global snakelike moves (lower chain).
These versions have been discussed elsewhere in detail \cite{DB91,Marcus,WBB95}
and we concentrate here on some salient features.

In the BFM each monomer occupies the eight corners
of the unit cube on a simple cubic lattice. The bonds between adjacent 
monomers can vary in length and direction, subject only to excluded volume 
constraints and entanglement restrictions.
Basically, due to the larger number of degrees of freedom 
--- $87$ angles between bonds are possible ---
the BFM does not suffer from the same ergodicity problems that
classical Monte Carlo schemes of self-avoiding random chains on 
simple cubic lattices are prone to \cite{Wall,Sokal}.
(We come back to this point in Sec.~\ref{subsec:reshisto}.) 
In the original BFM version only local jumps to nearest neighbor sites are allowed. 
The excluded volume then automatically forbids any intersection of bonds. 

\paragraph*{Slithering snake schemes: aSSA and qSSA.}
If snakelike moves are considered, the uncrossability of bonds has to be checked explicitly to 
avoid configurations which cannot be attained or unraveled by the original local scheme
\cite{WP96}.
The snake move along the existing backbone of a chain is equivalent
to removing an end monomer and connecting it to the other end of the chain,
leaving the lattice positions of the middle monomers unchanged. 
Therefore, density fluctuations occur only at chain ends if {\em no} local moves 
are allowed. 

The main focus of this study is the dynamics of the ``annealed SSA" (aSSA)
where all chains are allowed to move:
A chain and one of its ends are selected at random and a global move in 
a randomly chosen bond direction is attempted.
We compare this with the ``quenched SSA" (qSSA) of systems
where all chains but one are quenched, thus,
disallowing temporal density fluctuations 
(at chain ends or elsewhere).
We sample the dynamics of the remaining free snake and average
over all chains of one quenched equilibrium configuration.
Unless stated otherwise, in the following we always refer to aSSA properties.
Data from aSSA (qSSA) are generally indicated by open (full) symbols.

\paragraph*{Definition of a Monte Carlo step.}
Following Sokal \cite{Sokal} we have chosen to set the unit of the 
Monte Carlo time step (MCS) such that on average each monomer is attempted
once per cycle, irrespective of whether this attempt occurs 
through local or snake moves. 
If only the latter are allowed, every chain 
is on average visited once per MCS.
Our definition of the time scale differs from the one used traditionally
in the description of slithering snakes, which focuses on the computational time 
required to equilibrate samples. Indeed every snake move requires (essentially)
the same CPU time irrespective of $N$, since all monomers are moved simultaneously.
Hence, the computational relaxation time is proportional to $\taut/N$ where $\taut$ 
stands for the terminal relaxation time in our units.
%


\section{Parameters, configurations, measurements}
\label{sec:config}
This section presents various technical aspects of the simulation 
and introduces the important static and dynamical quantities for
the subsequent analysis. When a quantity is defined, we often refer to 
the figures where its dependence on $N$, $\phi$ or time is shown, 
without discussing the figure in detail, however. This discussion will 
be done in Sec.~\ref{sec:results}.

\paragraph*{Parameters.}
In this paper we study athermal systems (no additional potential parameters) 
containing monodisperse snakes (i.e., disallowing any local move) at volume
fractions $\phi$ ranging from dilute solutions to dense melts 
($\phi \approx 0.5$).  The volume fraction is defined by $\phi=8NN_\text{p}/
L^3$, where $N$, $N_\text{p}$ and $L$ are the chain length, the number of 
polymers in the system and the linear dimension of the simulation box,
respectively. In the present work, we primarily investigate the scaling of 
dynamical properties with mass $N$ 
(see, e.g., 
Figs.~\ref{figArateN},\ref{figDcurvN}--\ref{figTau},%
\ref{figg6alpha2},\ref{figDomegaN}).
Chain lengths ranging from $N=16$ up to $N=1024$ have been simulated.
This is sufficient to put theoretical predictions to the test.  
We use a periodic cubic lattice of linear size $L=128$.
At $\phi=0.5$ this corresponds to $131072$ monomers per simulation box.
Note that all length scales are given in units of the lattice spacing $a$
(i.e., $a=1$ in the following).

In fact, we report here the first results from a larger and still on-going study,
where in addition to $N$ and $\phi$ we have also varied 
the ratio $\omega$ of local to snake moves over several orders of magnitude
ranging continuously from purely local ($\omega=\infty$) to pure snake dynamics
($\omega=0$).  For the sake of comparison, the results of the local dynamics are 
included in some figures (Figs.~\ref{figAratephi},\ref{figDomegaN}).
At the end of this paper, we briefly comment on the influence that a finite value for
$\omega$ has on the dynamics (Fig.~\ref{figDomegaN}).
%

%

\paragraph*{Measurement of static properties.}
Obviously, the static properties of an ergodic system
do not depend on the dynamics of the algorithm.
This has been carefully checked for the annealed SSA.
(Numerically, it also appears to be true for the qSSA, even though ergodicity is 
less obvious in this case, since the environment of reference chains is frozen.
Note however that all initial qSSA chains are taken from an equilibrated
annealed ensemble.)
This allows us to use the precise characterization obtained in previous studies,
e.g., for the mean bond length $l$, the mean end-to-end distance 
$\Rend$
or the mean radius of gyration $\Rgyr$.
The first quantity gives the mean curvilinear distance along the
chain associated with every successful snake move, the second the typical
displacement $\Rend/N$ of the center of mass per snake move.
We recall that a polymer chain has a fractal dimension $1/\nu$ 
with $\nu$ being the Flory exponent: $\Rend \propto \Rgyr \sim N^{\nu}$.
The Flory exponent for dilute chains in a good solvent is $\nu=\nu_0 \approx 3/5$.
Above the overlap density $\phistar \approx N/\Rgyr^3 \sim N^{1-3\nu_0}$, 
the chains become Gaussian chains of blobs \cite{DG79} with $\nu=1/2$.
An important length scale for the subsequent discussion
is the mean distance between chain ends, $\dend = (4N/\phi)^{1/3}$,
which we will often refer to as ``mean end distance'' in the following.
Frequently, we use the ``effective bond length" \cite{DE86} 
$b(\phi) \equiv \lim_{N\rightarrow \infty} \Rend/N^{\nu=1/2}$
which is weakly density dependent.
This definition is chain length independent only for systems above $\phistar$.
In the dilute limit, we have $\Rend/N^{\nu_0}$, which yields $b_0 \approx 3$ for $N \gg 1$.
Some of these properties 
are summarized in Tables~\ref{tab:phi0.5} and \ref{tab:phi0.125}
for $\phi=0.5$ and $\phi=0.125$, respectively.
Note that $\Rgyr \gg \dend = 2N^{1/3}$ at $\phi=0.5$ for $N \ge 16$,
while for $\phi=0.125$ chains with $N < 64$ do not overlap, $\Rgyr < \dend$. 

\paragraph*{Various definitions of dynamic properties measured.}
As dynamical properties we have sampled various
spatial and curvilinear mean-square displacements (MSD) $g_i(t)$,
as well as associated times, diffusion coefficients 
and histograms. 
Following previous work \cite{PBHK91} we define the {\em spatial} MSD
$\gone(t)$ for the middle monomer (with monomer index $n=N/2$),
$\gtwo(t)$ for the motion of the middle monomer in the frame of the center of mass,
$\gthree(t)$ for the center of mass motion,
$\gfour(t)$ for the end monomer motion (averaged over $n=1$ or $n=N$)
and
$\gfive(t)$ for the motion of the end monomers in the center of mass frame.
Examples are given in Figs.~\ref{figmsddilute}--\ref{figg1melt}.
%
\label{DefMSD}
The MSD related to the {\em curvilinear} motion along the backbone is called
$\gsix(t) = \left< s(t)^2 \right>$ where $s$ is the curvilinear distance
traveled by the middle monomer $n=N/2$ in time $t$.
It is presented in Figs.~\ref{figg6melt}, \ref{figg1g3g6} and \ref{figg6scal}.
As we will see below, the scaling of $\gsix(t)$ and the associated histograms
(see Figs.~\ref{figTFPhisto},\ref{figg6histo},\ref{figg6alpha2})
are crucial for the understanding of the slithering snake dynamics.

The curvilinear and spatial diffusion coefficients,
presented in Tables~\ref{tab:phi0.5} and \ref{tab:phi0.125} and in 
Figs.~\ref{figDcurvN}, \ref{figDN} and \ref{figDomegaN},
are defined by
$\Dc \equiv \lim_{t\rightarrow \infty} \gsix(t)/2t$ and
$\Ds \equiv \lim_{t\rightarrow \infty} \gthree(t)/6t$.
\label{DefRelaxationTimes}
From the diffusion coefficients we introduce the time scales
$\taus \equiv {\Rgyr^2}/{\Ds}$ and
$\tauc \equiv {\left(N l \right)^2}/{(\pi^2\Dc)}$.
The prefactor in the last definition has been introduced for convenience
(see the discussion of Fig.~\ref{figTFPhisto} in Sec.~\ref{subsec:reshisto}).
Additional times can be introduced to characterize the different MSD's.
Following again Ref.~\cite{PBHK91}
the time scales \tauone, \tautwo, \tauthree \ and \taufour \
are defined by
$\gone(t \equiv \tauone)   = \Rgyr^2$,
$\gtwo(t \equiv \tautwo)   = 2\Rgyr^2/3$,
$\gtwo(t \equiv \tauthree) = \gthree(t=\tauthree)$ and
$\gfive(t \equiv \taufour) = \Rgyr^2$.
In addition to this we shall use
$\gone(t \equiv \tauend) = \dend^2$ and
$\gone(t \equiv \taud) = d^2$ 
where $d$ is an arbitrary distance.
To characterize the curvilinear motion of the chains
we also define the time $\tauseven$ for the diffusion of the middle monomer
over half of the chain's contour by $\gsix(t \equiv \tauseven) = (Nl/2)^2$. 
From the analysis it will turn out that two time scales, \tauone \ and \tauend,
are particularly important. They
have been included in Tables~\ref{tab:phi0.5} and \ref{tab:phi0.125}.


\paragraph*{Finite size effects.}
Finally, we note that the results obtained for $N=1024$ and 
$\phi=0.5$ have to be taken with some care. 
The configuration contains only $128$ chains and as $L \approx \Rend \approx 96$, 
we cannot rule out dynamical finite size effects --- 
especially for the quenched SSA.
Some very large motions have been observed due to chains taking advantage
of the image of their own correlation hole in the neighboring periodic boxes.
Unfortunately, a systematic study of the influence
of the box size $L$ is beyond our computational capabilities at 
present.


%
\section{Analytical expectations}
\label{sec:theory}
In the following paragraphes we summarize the relevant theoretical
expectations for the analysis of the simulation data.

\paragraph*{Mean-square displacements: general properties.}
If the chains do not become permanently trapped, one expects
free curvilinear and spatial diffusion, i.e.,
\begin{eqnarray}
\gsix(t) & \sim & 2 \Dc  t \label{eq:g6long} \\
\gone(t) = \gthree(t) = \gfour(t) & \sim & 6 \Ds t \;, \label{eq:g1g3g4long}
\end{eqnarray}
for times larger than the terminal relaxation time \taut \ and 
for distances larger than the radius of gyration \Rgyr. 
Furthermore, \gtwo \ and \gfive, both defined with respect 
to the center of mass of the chains, must yield a purely static quantity, 
the radius of gyration, 
\begin{eqnarray}
\gtwo(t) \approx \gfive(t)/4 &  \sim & \Rgyr^2 \;,\label{eq:g2g5long}
\end{eqnarray}
for very long times in any ergodic dynamical scheme \cite{PBHK91_JCP}. 

At short times the spatial displacement is directly expressible in
terms of the curvilinear displacement $s$ along a fractal
object characterized by the Flory exponent $\nu$.
For the MSD of the center of mass this suggests
\begin{equation}
N \gthree(t) \approx N \left< \left( 
\frac{\vec{R}_\text{e}}{N} \frac{s(t)}{l} \right)^2 \right> 
\stackrel{\text{MF}}{\approx} \frac{\Ree}{N} \frac{\gsix(t)}{l^2} \approx (b/l)^2 \gsix(t)
\label{eq:g3general}  
\end{equation}
where the penultimate step follows from a (seemlingly innocent) mean-field assumption.
The last step uses the fact that in the melt $\nu=1/2$. 
For the monomer displacements along the backbone a similar reasoning gives
\begin{equation}
\gone(t) =  \gtwo(t) \approx \gfour(t)=\gfive(t) \approx \left< R(s(t))^2 \right> 
\stackrel{\text{MF}}{\approx} b^2 \left(\gsix(t)/l^2 \right)^{\nu} 
\label{eq:g1g2g4g5general}
\end{equation}
where $\left< R(s)^2 \right>$ is the size of a piece of chain of
curvilinear length $s$. We have again made a mean-field approximation 
and assumed the fractal behavior of asymptotically long chains. 

\paragraph*{Uncorrelated reptation hypothesis.}
Until now everything is fairly general and is valid for both the URH and
the ARH. For uncorrelated snakes the motion along the backbone must 
be linear with respect to time. This fixes the short time predictions of 
the URH:
\begin{eqnarray}
\gsix(t)   \approx 2 \Dc t\,, &  & 
\gthree(t) \approx 6 \Ds t 
\label{eq:g6g3short}\\
\gone(t) = \gtwo(t) \approx \gfour(t)=\gfive(t) & \sim & t^{\nu}.
\label{eq:g1g2g4g5short}
\end{eqnarray}
Equation~(\ref{eq:g6g3short}) refers, strictly speaking, to the {\em approximately} 
free diffusion along the chain backbone. As our simulation does not deal with phantom 
chains, there must be a small 
correction even for very dilute solutions. 
(An isolated chain is not a phantom chain. It still interacts with itself.) 
At long times the MSD's must obey Eqs.~(\ref{eq:g6long}--\ref{eq:g2g5long}).

\paragraph*{Activated reptation hypothesis for a quenched molecular field.}
The description of the MSD's within the ARH is best outlined by 
considering chains in a molecular field that is quenched
at $t=0$ (qSSA). Hence, we start with an equilibrium ensemble of
reference chains. This implies that the chains initially sit in 
their correlation hole with relatively low conformational free energies.
To creep out of the local, well adapted environment a free energy penalty
must be paid on average. It is then reasonable to assume \cite{Semenov97,SR98}
(i) that fluctuations of the initial free energy may be neglected
(i.e., we consider a typical initial chain), 
(ii) that the curvilinear displacement $s$ may be taken as the 
{\em reaction coordinate} for the associated Kramers escape problem, and
(iii) that the same penalty $F(s)$ may be associated with
the possibly different conformations characterized by the same $s$
--- a simple, but less obvious mean-field approximation. 

To estimate the dependence of the energy penalty on $s$ it was argued in
Refs.~\cite{Semenov97,SR98} that $F(s) = f s^{\varphi}$ for $s \lesssim 
\splus(N)$ where $f$ is a constant.  The upper limit $\splus(N)$ 
depends on the relaxation (escape) mechanism.  Since we assumed the molecular 
field to be quenched, it seems natural to take $\splus \approx N \gg 1$: 
At distances larger than the radius of gyration, translational entropy 
must ultimately win over all other free energy contributions. 
To close the escape problem we still need to specify the physics for $s > 
\splus$.  Because of the gain of the translational entropy $F(s)$ should 
decrease logarithmically at large distances.  A crude, but physically 
sufficient approximation is to relax the penalty completely:
$F(s) \equiv 0$ for $s > \splus$ \cite{Semenov97}.%
\footnote{It is not sufficient to set $F(s)=F(\splus)$ for $s > \splus$
as this corresponds to a much too high return probability for the
one-dimensional Kramers problem into which 
we have cast the original three-dimensional model.}

The exponent $\varphi$ appears to be a matter of debate. 
In Refs.~\cite{Deutsch,SR98} a linear relationship was suggested ($\varphi=1$). 
This implies that the chemical potential for a curvilinear move is constant
and thus independent of the tail length $s$. On the other hand, it was 
suggested in Ref.~\cite{Semenov91,Semenov97} that the free energy penalty should 
result from fluctuations of the molecular field rather than from a fixed
chemical potential. The elaboration of this argument yields $\varphi=1/2$.

Given the free energy barrier one can calculate the mean-square displacements via
Eqs.~(\ref{eq:g3general},\ref{eq:g1g2g4g5general}) if one knows the probability
distribution $p(s,t)$.  In principle, $p(s,t)$ can be obtained from the solution
of the one-dimensional Smoluchowski equation. Analytically, this calculation has 
not been done yet,%
\footnote{We determined $p(s,t)$ in the simulations, see Fig.~\ref{figg6histo}.} 
but the qualitative behavior of the MSD's may be inferred from the following 
arguments.
At short times ($t \lesssim \tauminus$), we have $s \ll \splus$, and thus $F(s)/\kB T 
\ll 1$.  In this limit, one obviously recovers the free curvilinear diffusion of 
Eqs.~(\ref{eq:g6g3short}) and (\ref{eq:g1g2g4g5short}). If time increases, there 
should be an interval $\tauminus \ll t \ll \tauplus$ such that $\kB T \ll F(s) \ll
F(\splus)$.  Here, we expect that the chain is temporally localized 
in its correlation hole, which should lead to subdiffusive behavior due to 
intermittence of large scale displacements. The subdiffusive regime should extend up to 
a ``hopping time'' $\tauplus$. $\tauplus$ is the time taken by the chain to diffuse over
the curvilinear distance $\splus$ and thus to overcome the activation barrier. 
This implies $\tauplus \sim \splus^2 \exp [F(\splus)/\kB T]$. Only for $t \gg \tauplus$ 
the chain can diffuse freely.

\paragraph*{Activated reptation hypothesis for an annealed molecular field.}
Up to now, we considered the molecular field to be quenched. If all chains 
are allowed to move simultaneously, this annealing should reduce the activation
barrier.  For this case it has been suggested that 
$\splus(N) \sim \dend^{2} \sim N^{2/3}$ \cite{Deutsch,Semenov97,SR98}: The free
energy constraints can be relaxed by the reptation of other chains. 
For short times corresponding to $s \ll \splus$ the difference between a quenched 
or an annealed molecular field should be negligible. In this limit, the correlations 
between neighboring chains should be essentially of static nature. On the other 
hand, for $s \approx \splus$ the correlations are expected to be due to the 
cooperative rearrangements of several chains. Semenov and Rubinstein have 
proposed \cite{SR98} an intricate exchange mechanism which allows the 
relaxation of the free energy penalty between different snakes at $s \approx \splus$.
It is not obvious to us if one can still cast these processes in the simple
form of a one-dimensional Kramers escape problem.
In any case this should
only affect the probability distributions and MSD's around $s \approx \splus$. 

\paragraph*{Relaxation times and diffusion coefficients.}
The relative scaling of the terminal time $\taut$ and the diffusion coefficients
is imposed by Eq.~(\ref{eq:g3general}) which fixes the ratio of  
both diffusion coefficients: 
\begin{equation}
\taut \approx \frac{\left( N l \right)^2}{\Dc} \approx \frac{\Rend^2}{\Ds}\;.
\label{eq:taut}
\end{equation}
This ensures the matching of the long and short time behavior.
Hence, $\taut(N)$, $\Dc(N)$, $\Ds(N)$ are expected to contain the same information.
This should hold irrespective of whether the URH or the ARH applies.%
\footnote{In Sec.~\ref{subsec:restimes} it will be checked explicitly that this 
general relation still holds for the annealed SSA at different densities
(Fig.~\ref{figTauiTau1}). 
As we shall see, it fails, however, for the quenched SSA at high density
(Figs.~\ref{figmsdN128},\ref{figg1g3g6}).}

As we attempt to move all $N$ monomers of a snake collectively, the 
curvilinear diffusion coefficient should be independent of chain length for
uncorrelated snakes. Hence, it follows from Eq.~(\ref{eq:taut}) 
and from the URH that
\begin{equation}
\Dc/l^2 \approx m N^0 \mbox{ , } \Ds \approx m (\Rend/N)^2 \mbox{ , } 
\taut \approx \frac{N^2}{m} \;,
\label{eq:timesURH}
\end{equation}
$m$ being the $N$-independent mobility which, due to the URH, should be 
given by the acceptance rate, $m \approx A \sim N^0$.
Note that these equations differ by a factor of $N$ from the usual ones
characteristic of reptating chains within the URH \cite{DE86}. 
This is due to the fact that there, a factor $N$ is spent by the internal Rouse dynamics 
along the primitive path. It is precisely this missing factor which makes the SSA 
attractive in the first place to explore the dynamics of extremely long chains.

Within the ARH the curvilinear diffusion is fixed by the hopping time $\tauplus$
to reach a distance $\splus$ within the potential formed by the local
correlation hole, i.e., 
\begin{equation}
\Dc(N)/l^2 \approx \frac{(\splus/l)^2}{\tauplus} \approx 
m \exp\big[- F\big(\splus(N)\big)\big] = 
m \exp\big(- f \splus(N)^\varphi\big) \equiv \mplus \;,
\label{eq:DcARHgeneral}
\end{equation}
which in turn implies $\Ds(N)$ and $\taut(N)$ according to Eq.~(\ref{eq:taut}).
Equation~(\ref{eq:DcARHgeneral}) is obviously subject to the specific dependences
of $\splus(N)$ and $F(s)$ on their respective parameters.
We will try to estimate both $\tauplus(N)$ and $F(s)$ in 
Figs.~\ref{figg6scal} and \ref{figg6histo}.

%
\section{The dynamics of annealed and quenched snake fields}
\label{sec:results}

We begin our presentation of the slithering snake dynamics in 
Sec.~\ref{subsec:resArate} with the characterization of local 
phenomenological dynamical scales. 
Dilute and moderately dense systems are shown in Sec.~\ref{subsec:resdilute}
to be well described by the predictions of the URH.
Diffusion coefficients and characteristic time scales for 
a broad range of densities are presented in Sec.~\ref{subsec:restimes}.
We then focus on the dynamical properties of systems of dense solutions
and melts.
The various MSD's and their scaling are discussed in the next subsection.
By analyzing various displacement histograms possible ergodicity problems
are explicitly ruled out for the annealed SSA in Sec.~\ref{subsec:reshisto}.  
The histograms of the curvilinear displacements allow us to estimate
the free energy penality $F(s)$ mentioned above.

%
\subsection{Local phenomenological dynamical parameters}
\label{subsec:resArate}

\paragraph*{Acceptance rate.}
The simplest and statistically most accurate dynamical property sampled in any 
Monte Carlo simulation is the acceptance rate $A$ (= ratio of accepted to 
proposed moves).
It is included in the tables and plotted versus $N$ in Fig.~\ref{figArateN}
and versus $\phi$ in Fig.~\ref{figAratephi}.
We note {\em en passant} that the acceptance rates of the aSSA and the qSSA
are identical. The SSA acceptance rates are also completely independent of
chain length.
We recall that in contrast a weak, but systematic chain length dependence of $A$
is found for local dynamics due to the higher mobility of the chain ends compared to 
inner monomers (see, e.g., Fig.~10 of Ref.~\cite{PBHK91} and the bold line in 
Fig.~\ref{figArateN}).

As shown in Fig.~\ref{figAratephi}, $A$ is density independent below
$\phi \approx 0.01$, where it is also much larger for the SSA than for 
purely local dynamics.
This indicates that for these densities the blob size $\xi$ becomes sufficiently 
large to allow dilute SSA dynamics on distances smaller than $\xi$.
Above $\phi \approx 0.1$ density effects become pronounced and the
acceptance rate of the SSA decreases strongly. This decrease may be explained
as follows: one monomer is inserted in the 
SSA, i.e., 8 free lattice sites are needed, whereas 4 empty sites
suffice for local moves, in which a monomer is shifted from its
original position by one lattice constant. Hence, the observed smaller
acceptance rate for the SSA at large density ($\phi=0.5$) is expected.

\paragraph*{Local mobility.}
Obviously, it would be naive to believe that the acceptance rate of the Monte Carlo 
process necessarily sets the scale for the local mobility of the monomers.
There may be many allowed moves not contributing to the motion.
(They would also persist in a glass where the mobility is zero.)
In order to measure directly the monomeric mobility $m$ in the SSA
we exploit Eq.~(\ref{eq:g1g2g4g5short}) and compare the simulation data with
\begin{equation}
\gone(t) = b^2 \left( m t \right)^{\nu}.
\label{eq:mobilitydef}
\end{equation}
This equation provides a reasonable description of the MSD of the central monomer
at short times, as Fig.~\ref{figmsddilute} illustrates for a dilute solution and 
Fig.~\ref{figg1melt} for a dense melt.

For dilute systems it is appropriate to fit $\gone(t)$ using $\nu=\nu_0$ and the 
density independent effective bond length $b_0 \approx 3$. The same analysis can 
also be done in semidilute solutions for times shorter than the time required to 
cross the blob size so that the excluded volume interaction is not yet screened.  
This approach was also applied in Ref.~\cite{PBHK91} to extract $m$ for the
BFM in the case when only local moves are allowed. Our results for $m$ for both
the annealed and the quenched SSA are included in Fig.~\ref{figAratephi}.

However, Eq.~(\ref{eq:mobilitydef}) with $b=b_0$ and $\nu=\nu_0$ is not the only 
possible way to do the analysis.  For overlapping systems another choice is $b=b(\phi)$ and 
$\nu=1/2$.  This measures the effective mobility on distances of several blob sizes. 
Equation~(\ref{eq:mobilitydef}) then suggests that $\gone(t)$ should scale as
$\gone(t) \sim b(\phi)^2 t^{1/2}$ at short times, independent of chain length.  
Figure~\ref{figg1melt} confirms this expectation. If the mobility is now extracted  
from Fig.~\ref{figg1melt}, the differences between the new values and those obtained 
before (using $b=b_0$ and $\nu=\nu_0$) are negligible.

Here, the important point is that the local mobility is the same for all chain lengths 
at a given density.  
(Note that the mobilities obtained by Paul {\em et al.} \cite{PBHK91} 
for the local dynamics are also independent of chain length.)
Furthermore, Figs.~\ref{figmsddilute} and \ref{figg1melt} also 
show that $m$ is independent of whether the system is annealed (aSSA) or quenched
(qSSA).
For both dynamical schemes, the mobility rapidly decreases with density if 
$\phi > 0.1$.  This behavior is qualitatively the same as that of the acceptance rate. 
Similar results are obtained if Eq.~(\ref{eq:mobilitydef}) is fitted
to the MSD of the chain ends $\gfour(t)$ (not shown). For the SSA, the chain ends 
are more mobile than the central monomer by a factor of order one. 

In summary, the acceptance rate or the local mobility cannot explain 
the unexpected chain length dependence of the motion of strongly entangled chains
to be discussed in Secs.~\ref{subsec:restimes}--\ref{subsec:reshisto}.
%

\paragraph*{Effective mobility for the aSSA.}
A glance at Fig.~\ref{figg1melt} shows that in fact {\em two} fits
using Eq.~(\ref{eq:mobilitydef}) are possible for the annealed SSA 
once $\phi \gg \phistar$.
The rescaling $y=\gone(t)/b(\phi)^2t^{\nu=1/2}$ yields a data collapse
at very short times, $t \ll \tauminus$. From the resulting plateau the 
local mobility $\mminus$, shown in Fig.~\ref{figAratephi}, was extracted.
For larger times, however, $y$ decreases within the window $\tauminus 
\ll t \ll \tauplus(N)$. The existence of such a time window is obviously 
at variance with the uncorrelated reptation hypothesis.
For still larger times, $\tauplus \ll t \ll \taut$, the rescaled MSD
becomes approximatively horizontal. This allows us to define an
{\em effective} mobility $\mplus(N)$ (included in the tables). 
It reflects the complicated correlations in the second regime $\tauminus
\ll t \ll \tauplus$, which are responsible for its pronounced dependence
on $N$.
The importance of \mplus \ for the description of diffusion coefficients 
and relaxation times will be clarified in Sec.~\ref{subsec:restimes}.
Finally for $t > \taut$, Fig.~\ref{figg1melt} shows that $y \propto \sqrt{t}$, 
implying that the monomers diffuse freely.
%


%
\subsection{Dilute and weakly overlapping systems}
\label{subsec:resdilute}

Before analyzing the intricate physics at high chain overlap we briefly 
consider the simpler situation of dilute or weakly overlapping solutions.
These systems serve as a reference point for the following discussion
of concentrated solutions and melts.
That the predictions of the URH apply in the limit of high
dilution can be inferred from Figs.~\ref{figDcurvN}, \ref{figDN} and \ref{figTau}. 
The rescaling used for the vertical axes of the three figures is motivated by 
Eq.~(\ref{eq:timesURH})
and the fact that the local mobility $\mminus \approx A(\phi) \sim N^0$.
The figures confirm the validity of the URH by showing that
$\Dc \sim N^0$, $N\Ds \sim N^{2\nu_0-1}$ and $\taut \approx \tauone \sim N^2$.

Figure~\ref{figmsddilute} presents the spatial MSD's for a typical dilute 
or weakly overlapping configuration. We compare the measured MSD's with
the expected long time behavior (Eqs.~(\ref{eq:g1g3g4long},\ref{eq:g2g5long})) 
and the short time predictions of the URH (Eqs.~(\ref{eq:g6g3short},\ref{eq:g1g2g4g5short})
using $\nu=\nu_0$).
Note that the MSD of the center of mass $\gthree(t)$ is linear over 
several orders of magnitude, implying that corrections due
to excluded volume correlations of the snake with itself
are very small. The same applies to the
curvilinear motion $\gsix(t)$ (not shown). 
Incidentally, the excellent agreement confirms that our program is running properly
and that the URH is appropriate as long as chain interactions are weak.
The presented data are for the annealed SSA, but absolutely identical results have 
been obtained for qSSA, as expected. 

The above statements remain valid for all configurations and both 
dynamical schemes provided that $\phi \lesssim \phistar$. 
%

\subsection{Diffusion constants and time scales}
\label{subsec:restimes}

\paragraph*{Diffusion coefficients for the aSSA.}
The curvilinear and spatial diffusion coefficients, $\Dc$ and $N \Ds$,
are displayed in Figs.~\ref{figDcurvN} and \ref{figDN} not only for 
dilute solutions, but also for densities ranging up to the melt density
$\phi=0.5$. 
We plotted $\Dc/A$ and $\Ds N/A b^2$ {\em versus} chain length, since
we expected to find more or less horizontal curves 
(see Eq.~(\ref{eq:timesURH})).

Surprisingly, Eq.~(\ref{eq:timesURH}) breaks down with increasing chain
overlap.  For a given density the data points become progressively
curved in log--log coordinates.
In fact, \Dc \ and $N \Ds$ 
drop over more than two decades at $\phi=0.5$, as $N$ increases.
It is important to stress that the chain length dependence of \Dc \ and
$N\Ds$ increases gradually with chain overlap and that this effect is by 
no means a pathology of a lattice model at high density. 

The data for $\phi=0.5$ may be compared to a phenomenological power
law with exponent $\beta=4/3$. This power law
is merely a guide to the eye and a rough attempt to
characterize the slope in log--log coordinates for the available chain 
lengths.
However, the stretched exponential
$\Dc \approx N \Ds \approx \exp(-0.8 N^{1/3})$
motivated by Eq.~(\ref{eq:DcARHgeneral}) describes the data much better.
Our scaling is, in fact, in excellent agreement with the ARH and
the proposed energy barrier exponent $\varphi = 1/2$ \cite{Semenov97}.
The alternatively suggested value $\varphi=1$ \cite{Deutsch,SR98} 
is not compatible with the simulation data (at least for the chain 
lengths simulated up to now).

If, instead of the acceptance rate, we use $\mplus(N)$ to rescale the diffusion
coefficient, i.e., $N\Ds/\mplus b(\phi)^2$, we find a very good data collapse
(small symbols at the top of Fig.~\ref{figDN}).
A similar collapse is also possible for $\Dc/\mplus$ (not shown).
This shows that it is possible to absorb the
complicated physics at short times into one parameter, $\mplus(N)$.
Thus, the dynamics for $\tauplus \ll t \ll \taut$
is perfectly described in terms of uncorrelated reptation dynamics along the 
fractal chains. It remains to describe and possibly explain the dynamics
in the {second regime, $\tauminus \ll t \ll \tauplus$, of Fig.~\ref{figg1melt}.

\paragraph*{Diffusion coefficients for the qSSA.}
With increasing chain overlap the qSSA diffusion coefficients decrease 
even more rapidly with density and chain length than their aSSA counterparts.
In fact, while we have obtained most of the curvilinear diffusion coefficients
from our qSSA simulations, we have been unable to measure $\Ds$ for $\phi  > 0.125$. 
This surprising statement is obviously 
at variance with Eq.~(\ref{eq:taut}). It indicates that curvilinear and spatial
motions decouple for these high densities due to the localization 
of the snakes within their correlation hole in the quenched case. 
Furthermore, at high density \Dc \ is found to be a power law without any 
detectable curvature (Fig.~\ref{figDcurvN}). However, it is interesting to check 
whether Eq.~(\ref{eq:DcARHgeneral}) with $s_+(N) =N$
for the qSSA and $\varphi=1/2$ (for both schemes) yields a reasonable fit. 
This is indeed the case: $\Dc/A \approx 20 \exp(-0.15N^{1/2})$ is hardly
distinguishable from the indicated slope $\beta \approx 1.5$. 
Longer chains are obviously necessary to confirm the exponential behavior 
expected by the ARH.

\paragraph*{Relaxation times for the aSSA.}
In Figs.~\ref{figTauiTau1} and \ref{figTau} 
we display some of the characteristic times defined at the end of Sec.~\ref{sec:config}.
In the first figure we test Eq.~(\ref{eq:taut})
by plotting various ratios $\taui/\tauone$.
Apparently, the ratios are independent of chain length. We find
$\taus/4 \approx \tauc/5 \approx \tauone \approx \tautwo/0.85 \approx \tauthree/4.7 \approx
\taufour/0.3 \approx \tauseven/6$.
Only the data for $\phi=0.5$ have been included, but the same results prevail for 
all densities.  This demonstrates that 
all indicated times are proportional to one characteristic time scale,
the ``terminal time" \taut, containing the same information 
as the diffusion coefficients. That is, Eq.~(\ref{eq:taut}) 
still holds, although Eq.~(\ref{eq:timesURH}) fails.

Accordingly, Fig.~\ref{figTau} shows, e.g., that $\tauone$
is proportional to $N^2$ in the dilute limit, but increases like
$\tauone \approx N^2/\Dc \sim N^{2} \exp(0.8N^{1/3})$ in the melt. 
The figure also includes $\taud$ for $d=3$ and \tauend.
The first time indicates the distance $d$ up to which $\gone(t)$ remains
independent of $N$, and characterizes the extent of the first dynamical 
regime displayed in Figs.~\ref{figg1melt} and \ref{figg6melt}.
In fact, it strongly decreases with monomer density, and we expect it to
scale like the time needed to diffuse over the excluded volume blob $\xi$.
From the observed behavior of \Dc \ one anticipates for $\tauend$ the scaling 
$\taut \gg \tauend \approx s_{\text{end}}^2/\Dc \sim N^{2/3\nu} 
\exp(0.8N^{1/3})$, where $s_{\text{end}} \sim \dend^{1/\nu}$.
This is again consistent with the data presented in the figure.
%
%

\paragraph*{Relaxations times for the qSSA.}
Data points for $\tauone$ and $\tauseven$ at $\phi=0.5$ have 
been included for the qSSA in Fig.~\ref{figTau}.
Unfortunately, the range of $N$, which could be simulated, is insufficient to check
if indeed $\tauone \approx N^2 \exp(\text{const} N^{1/2})$.
The time scale \tauseven, measuring the curvilinear motion
over $N/2$ monomers, shows an effective power law which is
consistent with the corresponding diffusion coefficient
from Fig.~\ref{figDcurvN}:
$\tauseven \approx N^2/\Dc \sim N^{2+1.5}$.
Note that $\tauone$ and $\tauseven$ scale differently in
contrast to what was observed for the annealed SSA 
(Fig.~\ref{figTauiTau1}).
%

%


%
\subsection{Mean-square displacements of dense snakes}
\label{subsec:resmsd}

\paragraph*{Long time limit.}
Focusing on systems in the melt limit ($\phi=0.5$) we now discuss several 
MSD's and their scaling in detail (Figs.~\ref{figmsdmelt}--\ref{figg6scal}).
As a glance at the presented MSD's shows, the long time predictions 
(Eq.~(\ref{eq:g6long}--\ref{eq:g2g5long}))
are in very good agreement with the data for the aSSA, indicating that the chains
do not get trapped and that the simulation runs have been long enough.
This can be seen in Fig.~\ref{figmsdmelt} for the spatial MSD's
for $N=512$, in Fig.~\ref{figmsdN128} for both spatial and curvilinear MSD's
for $N=128$ and in Fig.~\ref{figg6melt} for the curvilinear MSD $\gsix(t)$.
%

The situation is more complex for the data of the qSSA
(included in Figs. \ref{figmsdN128}--\ref{figg6melt}).
Figure~\ref{figmsdN128} shows that
the spatial MSD's increase (approximately) logarithmically for the 
longest times we have been able to simulate.
This must be a transient: For very long times these MSD's
must either become constant, i.e., the chains become localized
on a length scale (presumably) given by the radius of gyration,
or ultimately, they diffuse freely in space.
So far, we have been able to show conclusively only the correctness of 
the second proposition for small chains ($N \le 32$).

At first sight surprisingly,  it is much less difficult to reach
the free {\em curvilinear} diffusion limit,
as may be seen from
Fig.~\ref{figmsdN128}, where $\gsix(t)$ and $N\gthree(t)$ are directly
compared for $N=128$, and from Fig.~\ref{figg6melt}, where the
$\gsix(t)$ for different masses are presented.
This indicates that for the qSSA the curvilinear and spatial
MSD's {\em decouple} for $t \gg \taucoup$ and $s \gg s_{\text{coup}}$ 
(see Fig.~\ref{figmsdN128}); although unable to move much further 
in space, the snakes manage to move relatively rapidly within the cage
of their correlation hole.
This is why we prefer to call this surprising behavior of dense
qSSA snakes (permanent or transient) {\em localization}
in contrast to a trivial local {\em blocking} of the snake moves
(on a scale given by the monomer density),
which must obviously occur at very high densities.

\paragraph*{Short time limit for the aSSA.}
While the MSD's are well described by the power laws of the URH
in the dilute limit (Fig.~\ref{figmsddilute}), this becomes strikingly 
different at larger
densities in the time window $\tauminus \ll t \ll \tauplus$. 
This is revealed by the direct comparison of 
dilute and dense systems with $N=512$, as given in
Figs.~\ref{figmsddilute} and \ref{figmsdmelt}. 
In this paper we attempt to characterize and to explain the 
features visible in the latter plot. 
%

We have tentatively described the subdiffusive behavior of the
aSSA data by means of an {\em effective} power law:
\begin{equation}
N\gthree(t) \approx \gone(t)^2 \approx \gsix(t) \sim t^{\alpha} 
\mbox{ with } \alpha \approx 1/2\;.
\label{eq:alpha}
\end{equation} 
This power law provides a remarkable fit for $\tauminus \ll t \ll \tauplus$
(compare Figs.~\ref{figg1melt},\ref{figmsdmelt},\ref{figg6melt}), 
even though some curvature is evident in log--log coordinates.
The alternative fit 
\begin{equation}
N\gthree(t) \approx \gone(t)^2 \approx \gsix(t) \sim \log(t)^{4\alpha}
\label{eq:logs}
\end{equation}
is less satisfactory for our aSSA data (fitting a smaller time interval),
as may be seen from Fig.~\ref{figmsdN128}.
(The exponents in the two previous equations have naturally
been chosen consistently with the general relationship expressed by
Eqs.~(\ref{eq:g3general},\ref{eq:g1g2g4g5general}) using $\nu=1/2$.)

\paragraph*{Short time limit for the qSSA.}
Equation~(\ref{eq:logs}), however, reasonably fits the qSSA data.
This may be seen from Figs.~\ref{figg1melt} and \ref{figmsdN128}.
While for the spatial MSD's this fit works for all times $t \gg \tauminus$,
$\gsix(t)$ becomes linear for $t \gg \tau_{+\text{q}}$, 
$t \gg \tau_{+\text{a}}$ with $\tau_{+\text{q}} \gg \tau_{+\text{a}}$
(compare aSSA and qSSA in Fig.~\ref{figg6melt} for $N=64,128$ and $1024$), 
where we altered the index to distinguish the different algorithms. 
Note that annealed SSA and quenched SSA data are identical for small times $t \ll \taufluc(N)$
(Figs.~\ref{figmsdN128},\ref{figg6melt}).
This is due to the low probability that a snake in the aSSA may 
interact with the density fluctuations created by neighboring snakes at short times.
Hence, the molecular fields probed behave in the same way.
As $\taufluc(N)$ increases strongly with $N$ (see below),
this suggests that the logarithmic behavior might be seen for
aSSA snakes as well, if we could simulate longer chains.
%
%

\paragraph*{Scaling of $\gone(t)$, $\gthree(t)$ versus $\gsix(t)$.}
In  Fig.~\ref{figg1g3g6} we verify now directly 
whether the time dependence of all spatial MSD's
is indeed fully encapsulated in terms of the curvilinear MSD $\gsix(t)$,
as suggested by both the URH and the ARH.
In the main panel of the figure we check this for the
aSSA data at $\phi=0.5$ by plotting $\gone(t)$ and $\gthree(t)$
{\em versus} $\gsix(t)$.
The MSD of the center of mass is linear with
regard to $\gsix$, as expected. This is in accord
with the previous finding $\Dc \approx N \Ds$ for Gaussian chains.
We find a similar data collapse for $\gone(\gsix)$ for $t \ll \taut$.
For sufficiently long chains, we clearly recover the anticipated
power law of Eq.~(\ref{eq:g1g2g4g5short}) for the diffusion on an object of 
inverse fractal dimension $\nu=1/2$ (dashed line in Fig.~\ref{figg1g3g6}).
(The departure from the slope $1/2$ for small $\gsix(t)$ is due to the
non-gaussian statistics for small distances along the chain.)
We conclude that the curvilinear and spatial diffusion are translated into each other,
as expected, and that the departure from the URH can be traced back to
the anomalous curvilinear diffusion depicted in Fig.~\ref{figg6melt}.

In the inset of Fig.~\ref{figg1g3g6}, we present $N\gthree(\gsix)$ for the qSSA
for two chain 
lengths and two densities. While for the lower density $\gthree(t)$ remains a (roughly)
linear function of $\gsix(t)$, this is only true for short times
at $\phi=0.5$, where the spatial displacements increase much more strongly.
Note that both densities appear to be independent of chain length.
This indicates a very weak mass dependence for $s_{\text{coup}}$. 

%
%
\paragraph*{Scaling with chain length.}
We now discuss in more detail the chain length dependence
of the subdiffusive behavior at fixed volume fraction $\phi=0.5$.
Since we have shown (Fig.~\ref{figg1g3g6}) that the spatial MSD's of the 
annealed SSA
scale in terms of $\gsix(t)$, we can concentrate on the $N$ dependence
of the latter quantity (Fig.~\ref{figg6melt}). 
The vertical axis is normalized to yield
the curvilinear diffusion coefficient \Dc \ as the plateau value for long times.
At short times $t \ll \tauminus \approx 50$, all curves merge.
Hence, \tauminus \ is chain length independent, but depends weakly on the
density, as may be shown by a similar plot containing data for different $\phi$.
For times beyond $\tauminus$ the dynamics slows down more and more until it reaches
the plateau of free diffusion at $\tau_{+\text{a}}(N)$ for aSSA and $\tau_{+\text{q}}(N)$ 
for qSSA, respectively.
Apparently, the correlations become more and more pronounced with increasing $N$.
%

We have directly computed the four times 
$\tau_{+\text{a}}$, $\tau_{+\text{q}}$, \taufluc \ and \taucoup \ 
and the related curvilinear displacements. 
Our measurements are not very precise (particularly, for the qSSA data better 
statistics is needed), and we only present some general trends here. 
A comparison of all chain lengths shows that 
\begin{eqnarray}
\taufluc(N) & \approx & \tau_{+\text{a}}(N) \approx \tauend(N)
\label{eq:tauflucaend}\\
\taucoup(N) & \approx & \tau_{+\text{q}}(N) \sim N^2 \;.
\label{eq:taucoupq}
\end{eqnarray} 
In other words, both pairs of time scales correspond to different
definitions of the same physical processes.
In the first case, these are the density fluctuations in the annealed scheme,
causing aSSA and qSSA to differ and the curvilinear motion to become 
eventually diffusive.
The second equation corresponds to the free curvilinear diffusion of the 
snake trapped in its quenched correlation hole. The power law 
indicated in Eq.~(\ref{eq:taucoupq}) is an estimate
motivated by the data given in Fig.~\ref{figTau} (diamonds). 
The last claim of Eq.~(\ref{eq:tauflucaend}), which is central to the
understanding of the observed correlations and is in line with
the ARH \cite{Deutsch,SR98}, is further tested in Fig.~\ref{figg6scal}.
The attempt $\tau_{+\text{a}} \approx  \tauone$ fails to scale the crossover
regime between the subdiffusive time window and the free curvilinear
diffusion at late times. (Note that there is always the time
scale $\tauminus \sim N^0$ for the free diffusion at very short times
and no scaling covering all regimes is possible.)
Instead, the alternative $\tau_{+\text{a}} \approx \tauend$ is successful.
The observed scaling confirms Eq.~(\ref{eq:tauflucaend}) and demonstrates
that the crossover from correlated to mean-field behavior is determined by \dend.
%

Finally, we note for the second, less important, relaxation process for the qSSA
that $s_{+\text{q}}$ increases very weakly with mass, $s^2_{+\text{q}} \sim N^{0.25}$.
(However, this does not contradict the scaling documented in the inset of 
Fig.~\ref{figg1g3g6}. Note also that this is again a rough estimate, 
but we definitely have to rule out the two ``natural" guesses $s_{+\text{q}} \sim N$ or 
$s_{+\text{q}} \sim N^0$.)
Interestingly, $s^2_{+\text{q}} \sim N^{0.25}$ and $\tau_{+\text{q}}(N) \sim N^2$ are 
compatible with the 
(very precisely measured) qSSA diffusion coefficient from Fig.~\ref{figDcurvN}, 
$\Dc \approx s^2_{+\text{q}}/\tau_{+\text{q}} \sim 1/N^{1.5}$, and, hence,
with the time scale $\tauseven \sim N^{3.5}$ shown in Fig.~\ref{figTau}.
While this is satisfactory, we presently cannot
explain the scaling behavior of $\tau_{+\text{q}}$ and $s_{+\text{q}}$. 

\subsection{Histograms and Non-Gaussianity}
\label{subsec:reshisto}

So far, we have discussed results obtained from the sampled
mean-square displacements and related diffusion constants
and time scales. We now turn to the additional information
contained in the histograms and higher moments of the {\em curvilinear} displacements. 
Merely results from the annealed field dynamics are discussed here,
since statistics is too poor for the qSSA.
  
An important, not only technical, question concerns the possibility 
that some chains or monomers become locally completely blocked.
We have paid much attention to this issue and have carefully 
checked that, in fact, all chains and monomers ultimately move arbitrarily 
far in space for the densities studied ($\phi \le 0.5$). 
Clearly, for some higher density this must eventually break down.
One may readily construct by hand chain configurations of the BFM,
which lead to blocked chains for any dynamical scheme under consideration,
even if local moves are allowed. 
These extremely rare points of the phase space also exist for our slithering 
snake scheme, although they are apparently excluded from our initial configurations
(which is the physically important point to make here).
These states cannot be accessed (by construction) and in this strict mathematical 
sense our dynamics is not ergodic. However, this is completely irrelevant for the
thermodynamic and dynamic behavior.%
\footnote{
In this context it is also of relevance that one obtains
the same dynamics if a small amount of local hopping moves,
$\omega \ll 1$, is included in the algorithm which allows
additional transitions between points of the phase space. 
See Fig.~\ref{figDomegaN}.
}

\paragraph*{Histogram of first passage time.}
Two checks are presented in the histograms given
in Figs.~\ref{figTFPhisto} and \ref{figg6histo}.
The first figure presents the histograms of first passage times of the 
central monomer over a curvilinear distance $s = l N/2$
for various chain lengths and densities.
At long enough times, typically $10\,\tausix$, where $\tausix = 
\langle t_{\text{fp}} \rangle$
is the mean first passage time, all chains have crept out of their initial tube.
If properly rescaled, the data collapse onto a master curve.
The distribution of the first passage times can be calculated
if free curvilinear diffusion is assumed.
It is a linear superposition of modes $p$ with an exponential time dependence 
$\exp(-p^2 t_{\text{fp}}/\tauc)$. For long times, this sum is dominated
by the lowest mode and, accordingly, we find an exponential decrease 
$\exp(-t_{\text{fp}}/\tauc)$ for all systems.
As expected, the mean first passage time scales like the terminal time,
$\tausix \approx 6 \tauone \approx 5 \tauc /6$,
and is of the same order as \tauseven \
which was defined as the time where $\gsix(t)$ corresponds to the
same curvilinear distance.

\paragraph*{Histogram of curvilinear displacements $s$.}
Figure~\ref{figg6histo} presents the histogram of $s(\Delta t)$
for $N=1024$ and $\phi=0.5$ for various time intervals, as indicated. 
We focus on the interesting time window $\tauminus \ll t \ll \tauplus \ll \taut$.
First of all we stress that the histograms have (very qualitatively) the same shape 
and fall off monotonously without any singularity at the origin or elsewhere,
i.e., no snake gets blocked at small $s$.
If no correlations were present, the rescaled histograms should all
collapse onto a single Gaussian. 
More importantly, however, Fig.~\ref{figg6histo} shows
that the different dynamical regimes correspond to (slightly)
different distributions. While at times $t \ll \tauminus$ and 
$t \gg \tauplus(N=1024) \approx 0.2\cdot 10^8$
the distributions are in fact Gaussian%
\footnote{For clarity, both regimes are not included in the figure. Note that Gaussian behavior
is only found in the first regime at much smaller densities than the one discussed
in Fig.~\ref{figg6histo}. A sufficiently large blob is required to obtain a Gaussian
distribution.},
they are essentially {\em exponential} in the subdiffusive regime 
where density fluctuations are important. 
The distribution is thus much {\em broader} than originally expected,
but all moments of the distribution exist. 
Although this is not an equilibrium distribution measuring directly the free energy as
a function of $s$,
the Smoluchowski equation suggests that $-\log(p(s))$ should be proportional $F(s)$.
We have therefore compared the two proposed exponents for the activation penality 
$\varphi=1$ (bold line) and $\varphi=1/2$ (dots) with the data.
Unfortunately at large $s$, the statistics becomes too poor to distinguish
between both exponents. More work is required to relate the form
and scaling of the observed histograms to the respective MSD.

\paragraph*{Non-Gaussian parameter.}
The non-gaussian parameter 
$\alpha_2 \equiv \frac{1}{3} \frac{\langle s^4 \rangle}{\langle s^2 \rangle^2} -1$
gives a compact means to describe the Gaussianity of these distributions as a function
of time. This is presented in Fig.~\ref{figg6alpha2}. 
We recall that $\alpha_2=0$ for
a Gaussian and $\alpha_2=1$ for an exponential distribution.
At short times $\alpha_2(t)$ decreases rapidly and is chain length independent.
The non-gaussianity at these times is due to local monomer interactions.
In agreement with the histograms shown in  Fig.~\ref{figg6histo}, 
$\alpha_2(t)$ {\em increases} strongly for larger times.
This shows again that correlations become more pronounced.
The effect increases strongly with chain length in agreement with 
the subdiffusive behavior in Fig.~\ref{figg6melt}. 
Note that $\alpha_2$ becomes even much larger than $1$ for our largest chain 
$N=1024$, reflecting the existence of a broad tail in the distribution shown in
Fig.~\ref{figg6histo}. We caution, however, that finite size effects
might contribute to this effect, as the end-to-end distance of this
configuration is comparable to the box-size. It is possible that
some snakes manage to find their way rapidly through the periodic box by
taking advantage of their own groove spanning the box. 
For times larger than $\tauplus(N)$ the non-gaussianity vanishes again
for all $N$, as it should. 
%

%
\section{Summary and outlook: snake versus local moves}
\label{sec:conclusion}

\paragraph*{Summary.}
In this paper we reported first numerical results on the dynamics of monodisperse
polymer chains moving according to the slithering snake algorithm SSA (no local 
relaxation pathways).  The simulations were done with the three-dimensional bond-fluctuation
model (BFM). We investigated two different dynamical schemes by focusing mainly on the 
polymer melts (high monomer densities):
Either we allow all snakes to move (aSSA), which permits fluctuations of the molecular field 
at the ends of neighboring chains, or all snakes but one 
are quenched (qSSA).
This work concentrates on the annealed algorithm, whereas the second mainly serves as a benchmark
to investigate the role of density fluctuations due to neighboring chains.
We studied the scaling of various spatial and curvilinear mean-square displacements (MSD),
diffusion coefficients and relaxation times with chain length. Furthermore, we also presented 
the distributions of the curvilinear motion.
%

%
%
Our results are broadly in agreement with the ``activated reptation" hypothesis (ARH)
\cite{Deutsch,SR98}.
For the aSSA, we find that at high chain overlap the terminal relaxation time $\taut$ 
increases much more strongly with chain length $N$ than possible for uncorrelated snakes
which are obtained for dilute and weakly overlapping chains.
In fact, a stretched exponential $\taut \approx N^2 \exp(0.8N^{1/3})$ fits
our data well. Similar and directly related (Eq.~(\ref{eq:taut})) behavior can be 
found for the spatial and curvilinear diffusion coefficients, $N\Ds \approx \Dc$, 
and other time scales, such as \tauend, which measures the time needed for the diffusion over 
the mean distance between chain ends \dend. The stretching exponent $1/3$ 
is in agreement with the free energy penalty $F(s) \sim s^{1/2}$ ($s$ being
the curvilinear displacement) estimated in Ref.~\cite{Semenov97},
but is incompatible with the more recent suggestion $F(s) \sim s$ of 
Ref.~\cite{SR98} (at least for the chain lengths accessible in the present simulation).

A more detailed study of different moments of the displacements shows that the
snake dynamics is only uncorrelated (exhibiting a free curvilinear diffusion 
with Gaussian distribution) for times smaller than $\tauminus(\phi)\sim N^0$, 
which is determined 
by the blob size $\xi$, and for times larger than $\tauplus \approx \tauend$ 
(shown by the scaling collaps of Fig.~\ref{figg6scal}), which is 
due to the density fluctuations present for distances larger than $\dend$.
They are responsible for an additional time scale (other than the terminal time) 
in the annealed algorithm, allowing the chains to relax the free energy penality 
caused by their motion in a locally rigid and frozen network of obstacles retaining 
them in their correlation hole.
This is supported by the finding that the MSD's of both the aSSA and the qSSA, 
which are strictly identical at small times, become different around \tauplus.

In the intermediate time interval the dynamics becomes increasingly subdiffusive
and non-gaussian (Figs.~\ref{figg6histo},\ref{figg6alpha2}), as the chain length
is increased for both aSSA and qSSA dynamics. The precise description of the MSD's in this 
regime is not evident for the masses we have been able to simulate and our study is
not fully conclusive. While power laws yield nominally better fits for aSSA,
logarithmically slow relaxation processes might also be appropriate.
The observed behavior for the quenched dynamics 
(Figs.~\ref{figmsdN128},\ref{figg6melt}), which could provide an ``envelope''
that the annealed dynamics might follow more closely with increasing chain length,
is indeed better described in this way. 
Logarithmic behavior is, however, not easily reconciled with (mean-field)
ARH approach which predicts an algebraic relationship.
More simulations are warranted to clarify this issue.  

The curvilinear displacement histograms decrease monotonously 
without any singularity at small $s$. In the subdiffusive intermediate time
they decrease essentially exponentially (Fig.~\ref{figg6histo}).
This suggests that the free energy penalty $F(s)$ is indeed of power law form. 
A direct measurement of the acceptance rate to perform a step $s \rightarrow s+1$ 
is currently underway using the quenched scheme which should allow a direct 
computation of the free energy cost for an escape attempt of order $s$.

In summary, we believe that our study demonstrates that the activated reptation approach 
is relevant for the description of concentrated solutions and melts of purely slithering 
snakes without local motion ($\omega=0$).

\paragraph*{The mapping onto a locally realistic model.}
It is important to recall that the slithering snake dynamics discussed in this paper 
is highly artificial, as no local rearrangements are allowed and density
fluctuations can only occur at chain ends. The idea is that the
snake corresponds to the primitive path of the reptation model and each 
monomer to a coarse-grained tube segment of an originally local model. 
The mapping of both levels of description is 
non-trivial.  It is not evident whether it can be performed at all without
an additional tuning parameter, such as the frequency of local hopping moves. 
%
%

%
The mapping might involve the density. It is possible that some of 
the high molecular densities, which we studied here, are in fact too large 
to correspond to realistic volume densities of an underlying 
microscopic model.
A characteristic density might therefore exist below which
even asymptotically long snakes remain uncorrelated.
At present, our computational results do not seem to favor
this option, as suggested, e.g., by Fig.~\ref{figDN}. However, it
remains a possibility which is currently pursued using
lower densities and much larger chains such that
$1 \gg \phi \gg \phistar(N)$. 
In any case, a more detailed analysis of the density dependence
is warranted to establish the scaling of 
$\tauminus(\phi)$ and $\tauplus(N,\phi)$. In other words, it has 
to be confirmed unambiguously 
that the first time scale is related to the blob size $\xi(\phi)$ and
the second one to the mean distance between chain ends $\dend(N,\phi)$.

\paragraph*{On the importance of local motion.}
Another issue is related to the fact 
that our slithering snakes do not have any mechanism to relax constraints 
via local pathways. It might well be that some local dynamical
degrees of freedom are necessary to {\em tune} a mesoscopic computation 
scheme such that it becomes consistent with real reptational dynamics.
A finite fraction of local moves should weaken the strength of the 
obstacles imposed by neighboring chains. This is indeed borne out by
preliminary studies presented in Fig.~\ref{figDomegaN},
where the rescaled spatial diffusion coefficient is plotted for various ratios
of local to snake moves $\omega$.
For short chains ($N < 64$) $N\Ds$ decreases monotonously
with increasing $\omega$, since local moves are less efficient in exploring the
phase space and the confinement is neglible. As the mass increases,
we find a striking {\em non-monotonous behavior}. The dynamics first becomes
more rapid, as local dense fluctuations become more relevant.
This effect apparently saturates at a frequency of order $10$, which 
causes the diffusion coefficient to decrease again strongly with increasing
$\omega$ (at fixed $N$). 
A finite $\omega$ generates a lateral tube of size $d_\text{l}(\omega,\phi) > a$, 
which should become identical to the tube diameter $d_\text{e}(\phi)$ of the
reptation model in the limit of very large $\omega$.
Recalling that $d_\text{e}(\phi)\approx \xi(\phi)$ we speculate 
that a successful mapping of a SSA scheme on a local scheme may require
the tuning of the ratio $d_\text{l}(\omega,\phi)/\xi(\phi) = \text{const} \approx 1$,
which implies a different $\omega$ for each density.
We are currently pursuing further simulations to clarify these questions. 

\paragraph*{A final word.}
It might well turn out that even in this larger modeling space
no $\omega(\phi)$ can be found, where $\Dc \approx N \Ds$ remains chain
length independent for large $N$.
We would regard this as a strong evidence for the correctness of the
ARH for a model with fully realistic local dynamics. 
We believe, however, that astronomically long chains are required to obtain 
measurable effects.
%


\begin{acknowledgments}
During the course of this work, we had valuable discussions and 
obtained insights from many colleagues, especially
from M.~M\"uller, A.~Johner, M.~Fuchs, A.~N. Semenov
and S.~Obukhov.
LM and JPW acknowledge computational support by IDRIS/France.
EL acknowledges support through a fellowship from the Max Planck
Institute for Polymer Research (Mainz, Germany).
\end{acknowledgments}


\newpage
\begin{table}[t]
\begin{tabular}{|l||c|c|c|c|c|c|c|c|}
\hline
$N$  & $l$   & $b(N)$& $A$     & \mplus & $N\Ds$  & $\Dc$  &$\tauone$& \tauend  \\ \hline
16   & 2.602 &   2.9 & 0.08433 & 0.0164 &  0.057  & 0.136  & 299     & 376      \\
32   & 2.603 &   3.0 & 0.08478 & 0.0123 &  0.040  & 0.085  & 1588    & 1257     \\
64   & 2.603 &   3.1 & 0.08502 & 0.0071 &  0.022  & 0.045  & 14527   & 5434     \\
128  & 2.603 &   3.1 & 0.08515 & 0.0035 &  0.010  & 0.021  & $1.3 \cdot 10^5$& 27826    \\
256  & 2.603 &   3.2 & 0.08506 & 0.0012 &  0.0037 & 0.0075 & $1.6 \cdot 10^6$& 192900   \\
512  & 2.602 &   3.2 & 0.08637 & 0.0003 &  0.0010 & 0.0021 & $1.8 \cdot 10^7$& $0.4\cdot 10^7$  \\
1024 & 2.603 &   3.0 & 0.08606 & 0.0001 &  0.0002 & 0.0005 & $2.5 \cdot 10^8$& $0.2\cdot 10^8$  \\
\hline
\end{tabular}
\vspace*{0.5cm}
\caption[]{Various static and dynamical properties for slithering snakes
with aSSA dynamics at volume fraction $\phi=0.5$ versus chain length $N$:
The mean bond length $l$,
the effective bond length $b(N)=\Rend/N^{1/2}$,
the acceptance rate $A$ (Fig.~\ref{figArateN}),
the effective mobility $\mplus(N)$ from $\gone(t)$ (Fig.~\ref{figg1melt}),
the spatial self-diffusion constant $\Ds = \lim_{t\rightarrow \infty} \gthree(t)/6t$
(Fig.~\ref{figDN}),
the curvilinear diffusion constant $\Dc = \lim_{t\rightarrow \infty} \gsix(t)/2t$
(Fig.~\ref{figDcurvN})
and, finally, the time scales \tauone  \ and \tauend \ obtained from 
$\gone(\tauone)=\Rgyr^2$ and $\gone(\tauend)=\dend^2$, respectively
(Fig.~\ref{figTau}).
The first of these times indicates the terminal time ($\tauone \approx \taut$),
the latter separates the strongly correlated anomalous 
chain diffusion from the free curvilinear diffusion at longer times
($\tauend \approx \tauplus$).
The mean bond length $l$ and the acceptance rate $A$ are essentially chain length
independent while the effective mobility $\mplus$ clearly is not.  
\label{tab:phi0.5}}
\end{table}

\begin{table}[t]
\begin{tabular}{|l||c|c|c|c|c|c|c|c|}
\hline
$N$  & $l$   & $b(N)$& $A$     & \mplus & $N\Ds$ & $\Dc$  &$\tauone$& \tauend  \\ \hline
16   & 2.689 &   3.4 & 0.62045 & 0.15   &  0.91  & 1.7    & 25      & 116      \\
32   & 2.690 &   3.5 & 0.62493 & 0.15   &  0.86  & 1.5    & 94      & 188      \\
64   & 2.691 &   3.7 & 0.62759 & 0.13   &  0.79  & 1.2    & 533     & 743      \\
128  & 2.692 &   3.8 & 0.62888 & 0.11   &  0.66  & 1.0    & 2246    & 1680     \\
256  & 2.692 &   3.9 & 0.62940 & 0.09   &  0.57  & 0.8    & 15142   & 6425     \\
512  & 2.692 &   3.9 & 0.63008 & 0.085  &  0.36  & 0.5    & 94401   & 20059    \\
1024 & 2.692 &   4.0 & 0.63006 & 0.055  &  0.28  & 0.3    & 780000  & 26329    \\
\hline
\end{tabular}
\vspace*{0.5cm}
\caption[]{Same properties as in Table~\ref{tab:phi0.5},
but for a lower volume fraction $\phi=0.125$.
Note that the effective bond length $b(N)=\Rend/N^{1/2}$
increases slightly with chain length: 
The chain overlap is weaker than at $\phi=0.5$ and larger
chains are needed to screen the excluded volume correlations.
For the relatively dilute systems $N \le 64$ 
the time scale $\tauend$ becomes larger than the terminal
relaxation time \tauone.
\label{tab:phi0.125}}
\end{table}
\clearpage


\begin{figure}[t]
\rsfig{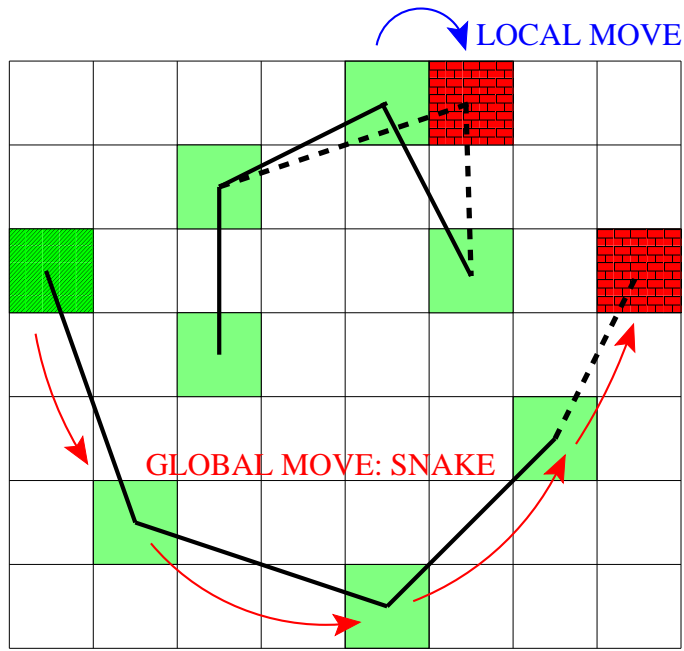}
\vspace*{0.8cm}
\caption[]{
Sketch of the bond-fluctuation model (BFM) with local (top chain)
and slithering snake moves (bottom chain).
We show a two dimensional projection of the three dimensional
Monte Carlo algorithm. Monomers (squares) are connected by 108 
possible bonds (bold lines) allowing for 87 angles between bonds
(in three dimensions).
In the original BFM, monomers make only {\em local} jumps in six spatial
directions, in (automatic) agreement with the imposed topological constraints.
%
In this paper we study the dynamics of the {\em slithering snake algorithm} (SSA). 
As depicted for the bottom chain, monomers move collectively along the chain in 
the SSA. Effectively, this amounts to removing a monomer
(the striped one on the left), connecting it to the other end of the chain
and leaving the middle monomers unchanged. Therefore, density fluctuations
and constraint release only occur at the chain ends.
%
%
%
\label{figsketchAlgo}}
\end{figure}

\begin{figure}
\rsfig{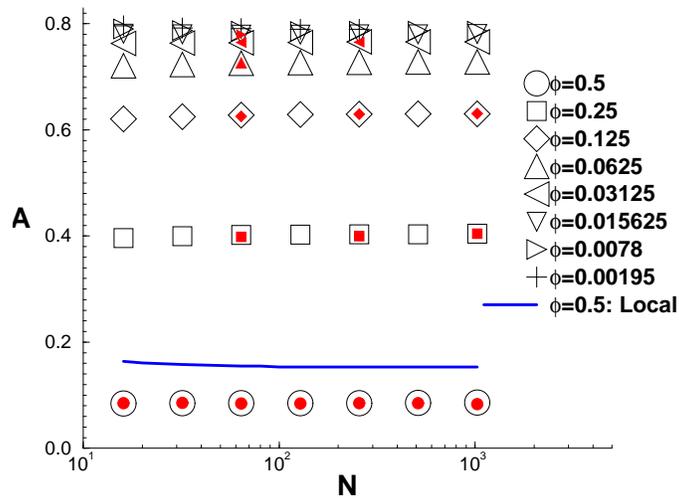}
\vspace*{0.4cm}
\caption[]{
Acceptance rate $A$ {\em versus} $N$ for various densities $\phi$, as indicated in the figure. 
The large open symbols correspond to the annealed SSA dynamics, while the small, filled
symbols indicate the quenched SSA data.
Both results are identical.
For the SSA the acceptance rate is independent of chain length. 
Contrary to that, the bold line indicates $A$ at $\phi=0.5$ for
the BFM with purely local moves. Here, a slight dependence on $N$ is found.
\label{figArateN}}
\end{figure}

\begin{figure}
\rsfig{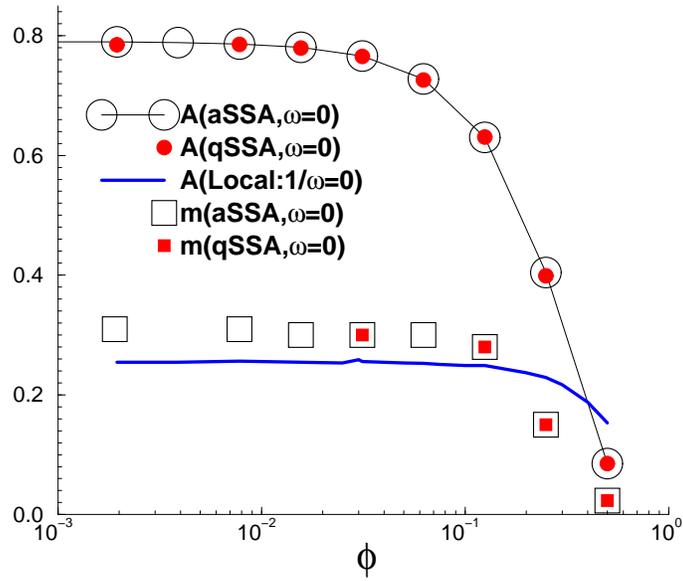} 
\vspace*{0.4cm}
\caption[]{
Acceptance rate $A$ (spheres) and local mobility $\mminus$ (squares) 
{\em versus} $\phi$ for pure slithering snake moves ($\omega=0$): annealed 
SSA (open symbols) and quenched SSA (small, filled symbols). 
Both $A$ and $\mminus$ are not affected by the freezing of the molecular field.
Also included are the acceptance rates from the BFM with local moves only 
($\omega =\infty$), extrapolated to infinite chain length: $\lim_{N\rightarrow \infty} 
A(N,\phi)$ (bold line).  For the SSA the acceptance rate and the local mobility
are independent of $N$.
\label{figAratephi}}
\end{figure}

\begin{figure}
\rsfig{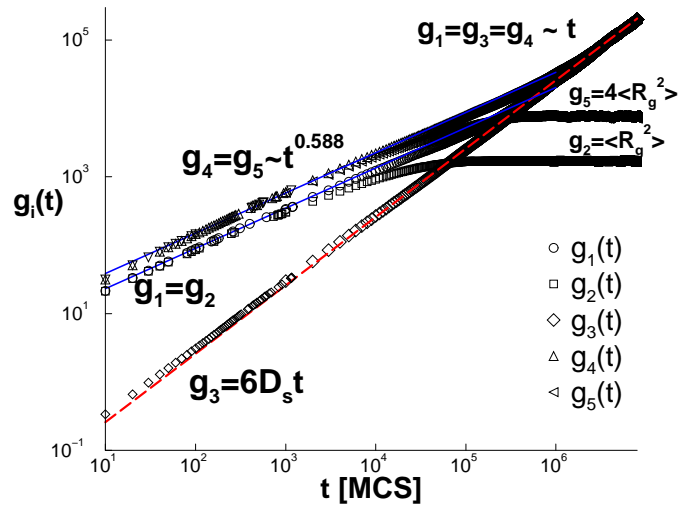} 
\vspace*{0.4cm}
\caption[]{
Different spatial mean-square displacements (MSD) $\gi(t)$
for a dilute system of chain length
$N=512$ at volume fraction $\phi=0.0156$. The MSD's are 
defined on page~\pageref{DefMSD}.
Only the data for the annealed SSA are given, as both dynamical schemes 
(aSSA and qSSA) yield the same results in the dilute limit.
At short times, $\gone(t)=\gtwo(t) \sim \gfour(t)=\gfive(t) \propto 
t^{\nu=0.588}$ (bold solid lines), 
as expected from Eq.~(\ref{eq:g1g2g4g5short}).
In good agreement with Eq.~(\ref{eq:g6g3short}), we also find $\gthree(t) \propto t/N$ over 
(almost) the entire time range. The straight (dashed) line for $\gthree(t)$ is a fit to 
the long time asymptotic limit. 
For large times $\gone(t)$ and $\gfour(t)$ must merge with $\gthree(t)$,
while $\gtwo(t) \rightarrow \Rgg$ and $\gfive(t) \rightarrow 4\Rgg$ converge
towards time independent static properties.
\label{figmsddilute}}
\end{figure}

\begin{figure}
\rsfig{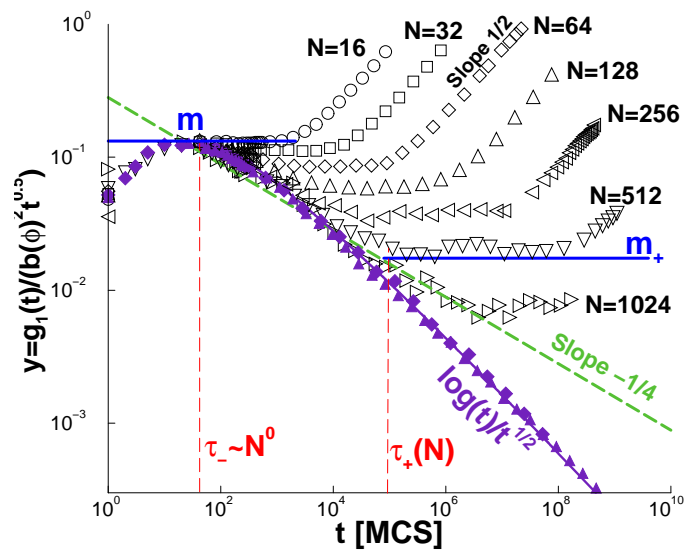} 
\vspace*{0.4cm}
\caption[]{
Estimation of local and effective mobilities, \mminus \ and \mplus.
The rescaled MSD $y=\gone(t)/(b(\phi)^2 t^{1/2})$ for the displacements of the central
monomer is shown at the melt density $\phi=0.5$. The mass $N$ varies, as indicated.
Open symbols refer to the annealed SSA.
We also included the qSSA results for $N=64$ and $N=128$ (filled symbols). 
For comparison,
we show a power law with exponent $-1/4$ (dashed line)
and $\gone(t) \approx 3.5 \log(t) - 3.8$ (bold line) which fits well the subdiffusive
behavior of the qSSA. 
From the two horizontal envelopes of the curves we estimate the 
short time mobility $\mminus(\phi) \sim N^0$ and the effective mobility of the snake dynamics
$\mplus(N,\phi)$. 
(The latter quantity is not visible for qSSA, as a much longer 
simulation run would have been required.) 
As can be seen from the figure, $\mplus$ strongly depends on chain length.
\label{figg1melt}}
\end{figure}

\begin{figure}
\rsfig{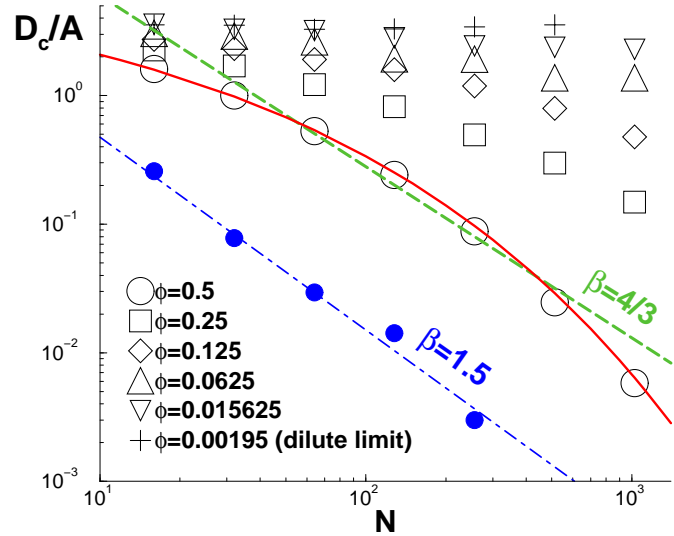} 
\vspace*{0.4cm}
\caption[]{
Curvilinear diffusion coefficient $\Dc/A$ {\em versus} $N$ for various volume fractions $\phi$ 
($A=$ acceptance rate).
The open symbols show the results for the annealed SSA, the filled symbols for the
quenched SSA.
Only in the dilute limit $\Dc/A \sim N^0$, as expected.
Although the effect strongly increases with $\phi$,
we find deviations from this expectation at all densities 
for large chain overlap $\phi/\phistar(N) \gg 1$.
The effective power law with exponent $\beta=4/3$ roughly characterizes the chain length 
dependence at $\phi=0.5$ for the aSSA dynamics. 
A better fit, which properly takes into account 
the curvature of the data, is given by the stretched exponential
$\exp(-0.8 N^{1/3})$ (bold line).
The qSSA data at $\phi=0.5$ are well fitted by an effective power law 
($\beta \approx 1.5$) without any noticeable curvature (dashed-dotted line). 
\label{figDcurvN}}
\end{figure}

\begin{figure}
\rsfig{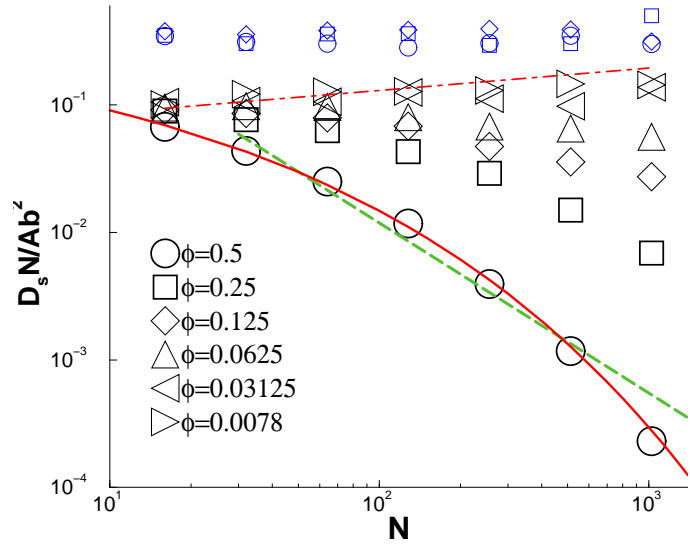} 
\vspace*{0.4cm}
\caption[]{
Spatial diffusion coefficient $\Ds=\lim_{t\rightarrow \infty} \gthree(t)/6t$
{\em versus} chain length $N$ for different volume fractions $\phi$,
as indicated in the figure (annealed SSA only).
The large symbols correspond to the rescaled value $N\Ds/ A b(\phi)^2$
where we use the acceptance rate $A(\phi)\sim N^0$ to characterize
the local mobility. 
The small symbols (top) have been obtained by
using the effective mobility $\mplus(N,\phi)$,
obtained from $\gone(t)$ (Fig.~\protect\ref{figg1melt}), as a measure of the
local mobility instead of $A$. 
The dash-dotted line corresponds to the prediction in the dilute limit, 
$\beta=2 \nu_0-1 \approx 0.176$. 
The data at $\phi=0.5$ are compared to the
effective power law with exponent $\beta=-4/3$ (dashed line)
and the stretched exponential $\exp(-0.8N^{1/3})$ (bold line).
\label{figDN}}
\end{figure}

\begin{figure}
\rsfig{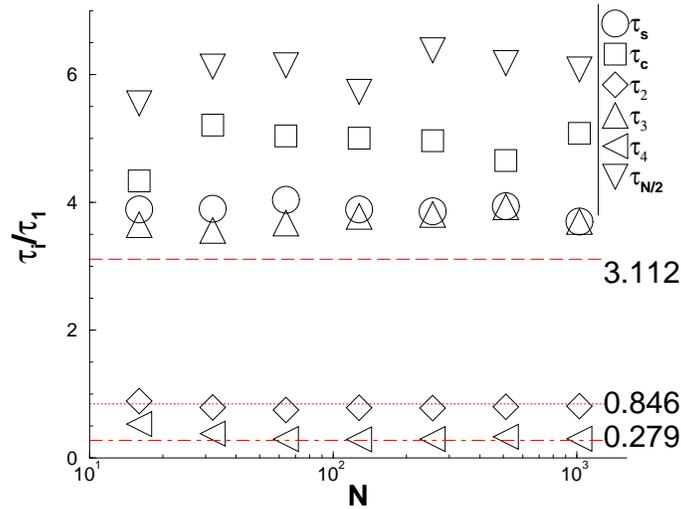} 
\vspace*{0.4cm}
\caption[]{
Rescaled times $\taui/\tauone$ (defined on page~\pageref{DefRelaxationTimes})
at $\phi=0.5$ (aSSA only). 
Similar behavior is also found for other densities.
This confirms that all these times are proportional to one characteristic time scale 
--- the ``terminal time" \taut \ ---
containing the same information as the diffusion coefficients.
The horizontal lines are comparisons with the Rouse model where $\taufour/\tauone \approx 0.279$,
$\tautwo/\tauone \approx 0.846$ and $\tauthree/\tauone \approx 3.112$.
Here, we did not include $\tauend$, as its scaling is different
(see Fig.~\protect\ref{figTau}).
\label{figTauiTau1}}
\end{figure}

\begin{figure}
\rsfig{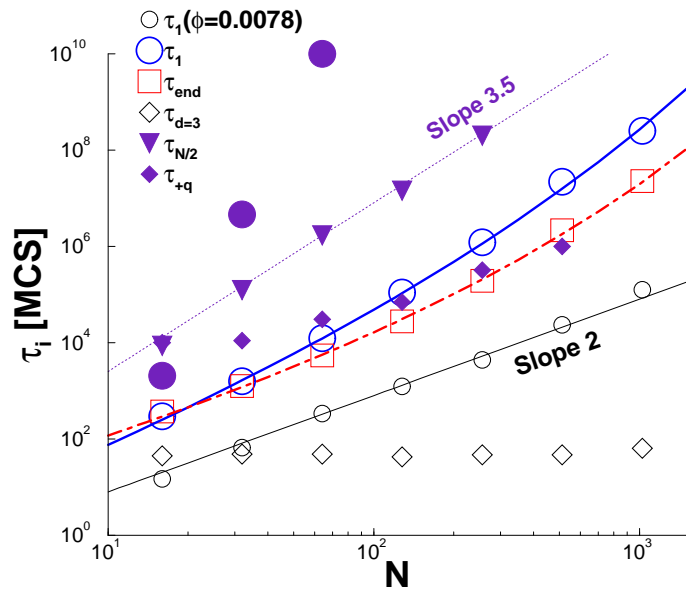} 
\vspace*{0.4cm}
\caption[]{
Unscaled times {\em versus} $N$. The open symbols present the results of the 
annealed SSA, the filled symbols of the quenched SSA. 
For the aSSA dynamics \tauone, \tauend \ and $\tau_{d=3}$ 
(defined on page~\pageref{DefRelaxationTimes}) are shown.  
In the dilute limit ($\phi=0.0078$)
we find $\tauone \sim N^2$, as expected (small spheres, thin line).
All other data are for $\phi=0.5$.
The stretched exponentials are motivated by the
ARH. They provide a good fit for both
$\tauone \approx N^2 \exp(0.8N^{1/3})$ (bold line) and 
$\tauend \approx N^{2/3\nu} \exp(0.8N^{1/3})$ (dash-dotted line) at high densities.
For the qSSA dynamics the filled spheres, triangles and diamonds correspond to 
$\tauone$, $\tauseven$ (also defined on page~\pageref{DefRelaxationTimes})
and $\tau_{+\text{q}}$ at $\phi=0.5$.  
$\tau_{+\text{q}}$ is the crossover time to free curvilinear 
diffusion for the qSSA (see Fig.~\ref{figmsdmelt} for its definition in the aSSA case).
As expected, the relaxation times for the qSSA increase much faster than for the
aSSA. An effective power law $\tauseven \sim N^{3.5}$ is found
in agreement with the corresponding diffusion coefficient $\Dc$ 
(see Fig.~\ref{figDcurvN}).
The crossover to free curvilinear diffusion for the qSSA is 
roughly characterized by $\tau_{+\text{q}} \sim N^2$ without noticeable curvature.
\label{figTau}}
\end{figure}

\begin{figure}
\rsfig{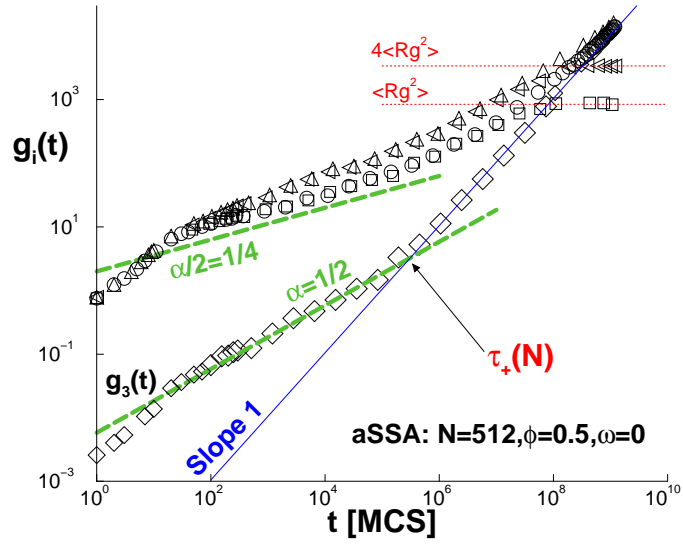} 
\vspace*{0.4cm}
\caption[]{
Spatial MSD's for a dense system of chain length $N=512$ at volume fraction $\phi=0.5$ 
for the annealed SSA (pure slithering snake dynamics, i.e., $\omega=0$). The symbols are 
the same as in Fig.~\ref{figmsddilute}.
%
While the behavior at large times is the same as in the previous figure, much slower
dynamics and strong curvature are found at short times. 
This is a typical result for strongly overlapping chains
(either large $\phi$ or large $N$). 
According to Eq.~(\ref{eq:alpha}) we compare 
$\gone(t)=\gtwo(t) \approx \gfour(t)=\gfive(t)$ and $\gthree(t)$ with 
power laws characterized by an effective exponent $\alpha=1/2$.
The time scale $\tauplus$ characterizes the crossover between
subdiffusive and free (linear slope) center of mass motion.
Its scaling with $N$ is discussed in Fig.~\ref{figg6scal}.
\label{figmsdmelt}}
\end{figure}

\begin{figure}
\rsfig{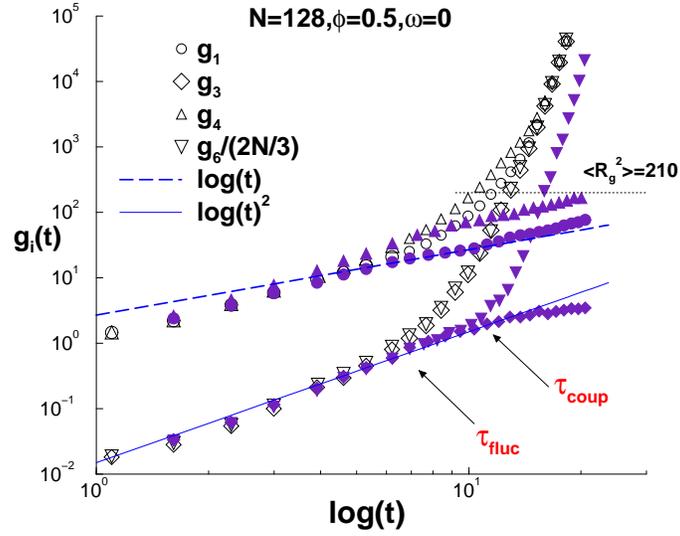} 
\vspace*{0.4cm}
\caption[]{
Double logarithmic plot of spatial and curvilinear MSD's for the annealed SSA 
(open symbols) and the quenched SSA (filled symbols) 
for $N=128$, $\phi=0.5$ {\em versus} time. 
As suggested by Eq.~(\ref{eq:g3general}), we rescaled the curvilinear MSD 
$\gsix(t)$ by $Nl^2/b^2 \approx 2N/3$ to compare it with $\gthree(t)$.
For the aSSA spatial and curvilinear motion are coupled for all times:
$\gsix(t) \sim N \gthree(t)$.
For the qSSA
the run is too short to reach the spatial free diffusion limit for $N=128$.
However, aSSA and qSSA dynamics are identical for small times $t \ll \taufluc
\approx \tauend$ (see Eq.~(\ref{eq:tauflucaend})).
Roughly logarithmic behavior is found for the qSSA dynamics, as indicated by the two
fits $\gone(t) \sim \log(t)$ and $N\gthree(t) \approx \gsix(t) \sim \log(t)^2$.
The time $\taufluc$ denotes the time where the density fluctuations of the
annealed field become relevant,
$\taucoup$ the breakdown of the coupling of curvilinear and spatial displacements for 
the qSSA.
\label{figmsdN128}}
\end{figure}

\begin{figure}
\rsfig{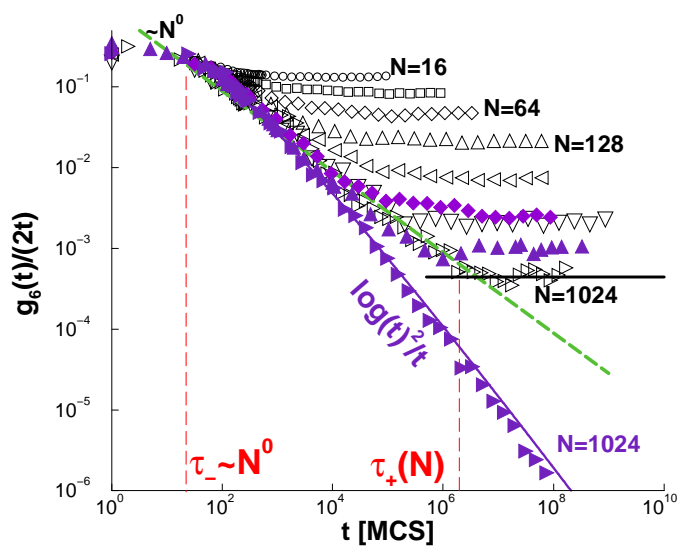} 
\vspace*{0.4cm}
\caption[]{
Rescaled MSD of the curvilinear diffusion of the central monomer $y=\gsix(t)/(2t)$.
The MSD is normalized to yield the curvilinear diffusion coefficient \Dc \ as a 
plateau value.
The annealed SSA data (open symbols) comprise all $N$  
for the density $\phi=0.5$. For comparison, we also included the qSSA
results for $N=64, 128, 1024$ (filled symbols).
At short times $t \ll \tauminus \approx 50$, all curves merge.
For larger times the dynamics progressively slows down as a function of chain length.
Only for times $t \gg \tauplus(N)$ (indicated here for $N=1024$) we obtain the expected 
free curvilinear diffusion.
The intermediate decrease for the aSSA dynamics
is very roughly fitted by the power law
$\alpha -1 \approx -1/2$ (dashed line).
The qSSA data for $N=1024$ are well fitted by
the logarithmic relation $\gsix(t)^{1/2} \approx \log(t)$ (bold line).
This suggests that for much longer chains even the aSSA dynamics might 
approach this logarithmic and chain length independent envelope.
%
\label{figg6melt}}
\end{figure}

\begin{figure}
\rsfig{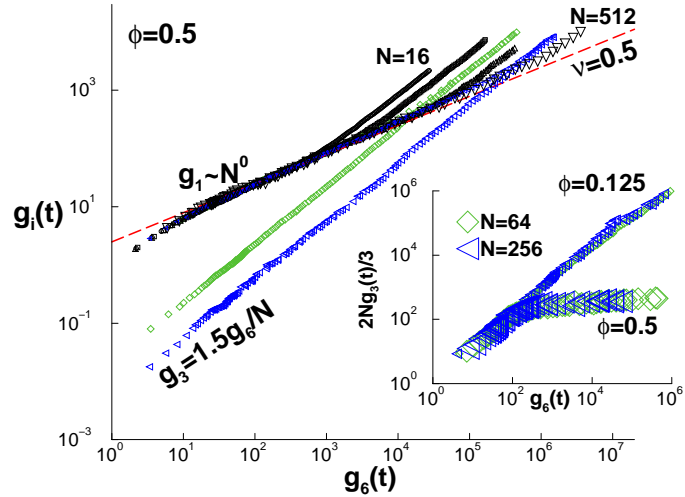} 
\vspace*{0.4cm}
\caption[]{
Relative scaling of the spatial and curvilinear displacements
for the {\bf (a)} aSSA (main figure) and {\bf (b)} qSSA (inset) dynamics.
The MSD's are defined on page~\pageref{DefMSD}.
{\bf (a)}  
The spatial MSD's $\gone(t)$ and $\gthree(t)$ are plotted {\em versus} the curvilinear 
MSD $\gsix(t)$ for various chain lengths, as indicated ($\phi=0.5$).
As $\gthree(t)$ is strictly linear with respect to $\gsix(t)$ for all $t$ and $N$,
we only included the data for $N=64$ and $N=256$ for the sake of clarity.
As expected, we find 
$\gone(t) \approx b^2 (\gsix(t)/l^2)^{\nu=1/2} \propto N^0$ for
$t \ll \taut$ (dashed line labeled by $\nu=0.5$).
Hence, the anticipated relations between curvilinear and spatial diffusion are
verified for the aSSA dynamics and the anomalous dynamics must be fully 
encapsulated in terms of the curvilinear displacement $\gsix(t)$.
{\bf (b)} Rescaled spatial MSD $2N\gthree(t)/3$ {\em versus} $\gsix(t)$.
Only at short times and for densities $\phi \le 0.125$ we find both
quantities to be equal. The curvilinear motion {\em decouples} from
the spatial motion for larger densities, where the snakes become
{\em localized} within the correlation hole. 
\label{figg1g3g6}}
\end{figure}

\begin{figure}
\rsfig{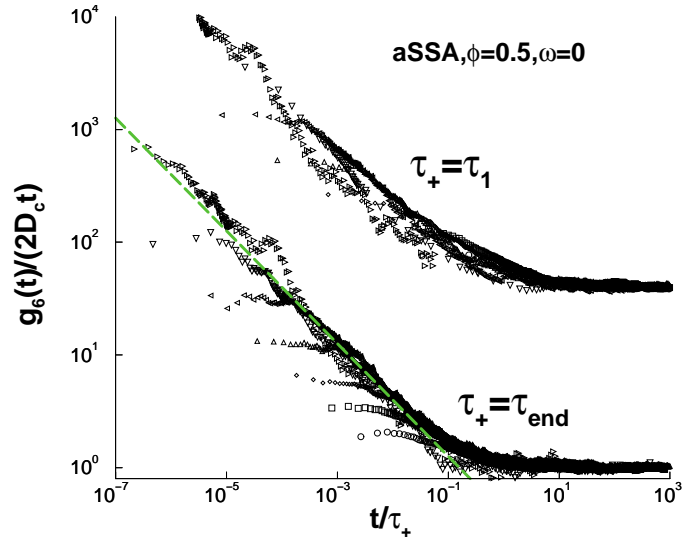} 
\vspace*{0.4cm}
\caption[]{
Scaling attempts for $\gsix(t)$ in order
to determine the time scale at which the curvilinear dynamics
(for the aSSA) becomes freely diffusive.
The natural attempt $\tauplus = \tauone$ (shifted for clarity)
fails. In contrast, the scaling $\tauplus = \tauend$
is successful, demonstrating that there is an additional characteristic time
related to the density of chain ends.
The power law with exponent $\alpha \approx 1/2$
for the envelope is indicated (dashed line). 
\label{figg6scal}}
\end{figure}

\begin{figure}
\rsfig{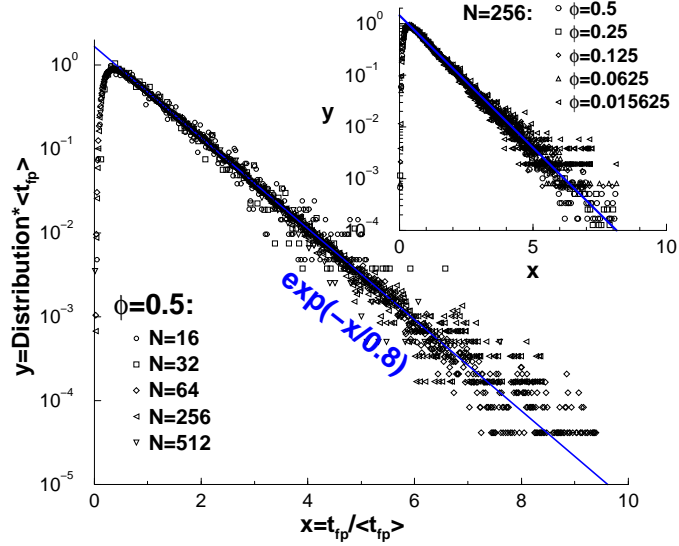} 
\vspace*{0.4cm}
\caption[]{
Distribution of the first passage time $t_{\text{fp}}$ required to reach 
a curvilinear distance $Nl/2$.
We consider aSSA systems at $\phi=0.5$ (main figure)
and at chain length $N=256$ (inset).
Symbols are indicated in the figure.
The data points of all systems $(\phi,N)$ collapse on the same master curve
if rescaled by the mean first passage time $\tausix \equiv \langle t_{\text{fp}} \rangle$.
At large times the histograms decrease exponentially according to $\exp(-t/\tau_\text{c})$, 
as expected.
\label{figTFPhisto}}
\end{figure}

\begin{figure}
\rsfig{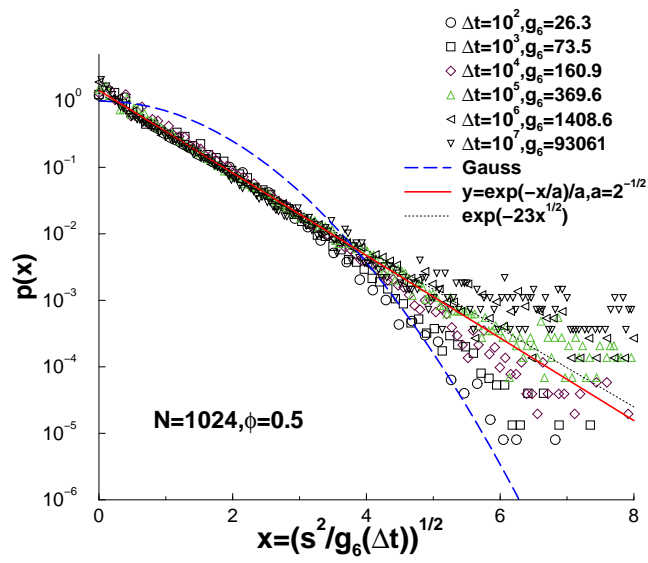} 
\vspace*{0.4cm}
\caption[]{
Normalized histograms of the curvilinear motion 
(aSSA data only) for $N=1024$, $\phi=0.5$ and
for different time intervals $\Delta t$, as indicated in the figure.
The histograms are rescaled in such a way that all data should
collapse if they were self-similar for all times.
We show that the histograms are approximately
exponential within the time interval 
$\tauminus \approx 50 \ll t \ll \tauplus=\tauend(N=1024) \approx 0.2\cdot 10^8$.
The histograms become Gaussian again at much longer times.
\label{figg6histo}} 
\end{figure}

\begin{figure}
\rsfig{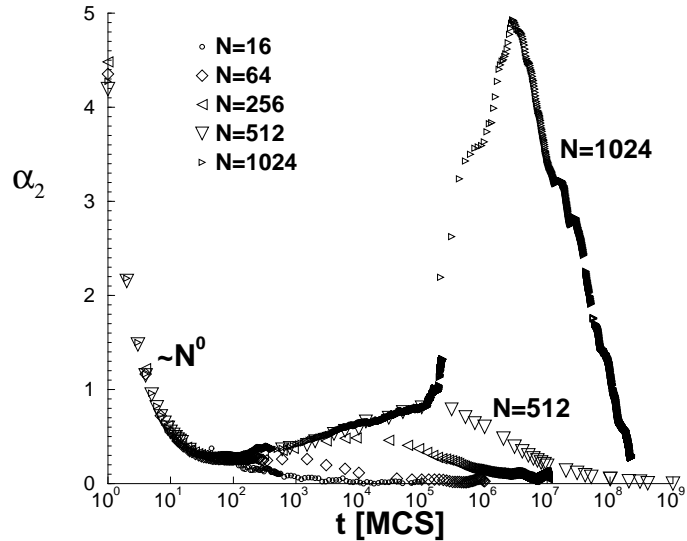} 
\vspace*{0.4cm}
\caption[]{
Non-gaussian parameter 
$\alpha_2 \equiv
\frac{1}{3} \frac{\langle s^4 \rangle}{\langle s^2 \rangle^2}-1$
for the curvilinear displacement $s$ of the central monomer 
{\em versus} time for $\phi=0.5$ and chain lengths as indicated in the figure.
We find that $\alpha_2$ becomes non-monotonous for 
$\tauminus \ll t \ll \tauplus$.
Unfortunately, the statistics is poor for large $N$
and some binning is necessary. 
Only the aSSA data are shown, the statistics is not sufficient for the qSSA.
\label{figg6alpha2}}
\end{figure}

\begin{figure}
\rsfig{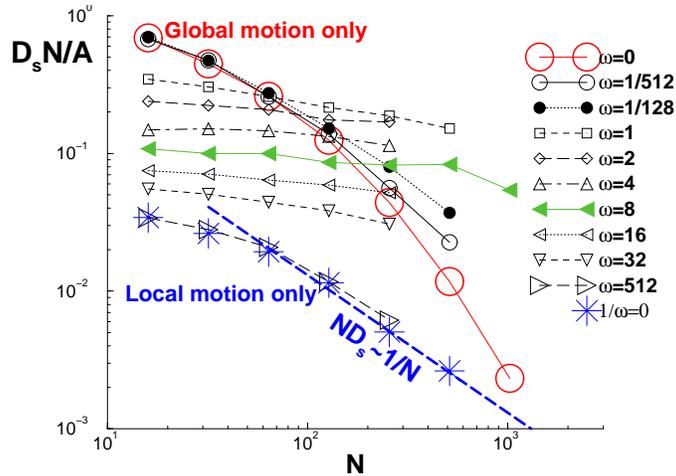} 
\vspace*{0.4cm}
\caption[]{
Spatial diffusion coefficient $N\Ds$ {\em versus} $N$ for different $\omega$
as indicated in the figure (aSSA only): 
$\omega=0$ corresponds to pure slithering snake dynamics, $1/\omega=0$ (large stars) 
to the pure local dynamics.
The data for $\omega \ll 1$ are very similar to the pure slithering snake limit.
For $\omega \approx 8$ Eq.~(\ref{eq:timesURH}) is approximately satisfied.
This may define a reasonable ``starting point"
for future SSA simulations corresponding to the classical
URH reptation approach. 
For the local dynamics the power law prediction $\Ds \propto 1/N^2$
(dashed line) from classical reptation theory
is indicated. The data for $\omega =512$ approximate the local dynamics limit 
rather well for the chain lengths we have been able to simulate.  
\label{figDomegaN}}
\end{figure}

\end{document}